\PassOptionsToPackage{colorlinks=true,linkcolor=blue,filecolor=black,urlcolor=blue,citecolor=magenta}{hyperref}
\documentclass[11pt]{article}
\usepackage{jheppub}
\usepackage{amsmath,amssymb}
\usepackage{mathrsfs}
\usepackage{amsfonts}
\usepackage{amsthm}
\usepackage[table]{xcolor}
\usepackage{caption,subcaption}
\usepackage{tikz}
\usetikzlibrary{decorations.markings}
\usetikzlibrary{patterns}
\usepackage{color, soul}
\usepackage{slashed}
\usepackage{arydshln}

\DeclareMathOperator{\arccosh}{arccosh}
\DeclareMathOperator{\arcsinh}{arcsinh}

\numberwithin{equation}{section}

\preprint{QMUL-PH-23-34}

\title{ 
An eikonal-inspired approach to the \\ 
gravitational scattering  waveform
}
\author[a]{Alessandro Georgoudis,}
\author[b]{Carlo Heissenberg,}
\author[b]{Rodolfo Russo}

\affiliation[a]{Centre for Theoretical Physics, Department of Physics and Astronomy, \\ Queen Mary University of London, Mile End Road, London E1 4NS, United Kingdom}
\affiliation[b]{
	School of Mathematical Sciences, Queen Mary University of London, \\ Mile End Road, London, E1 4NS, United Kingdom
}
\emailAdd{a.georgoudis@qmul.ac.uk,c.heissenberg@qmul.ac.uk,r.russo@qmul.ac.uk}
\abstract{
  We revisit the amplitude-based derivation of  gravitational waveform for the scattering of two scalar black holes at subleading post-Minkowskian (PM) order. We take an eikonal-inspired approach to the two-massive-particle cut needed in the KMOC framework, as highlighted in~\cite{Caron-Huot:2023vxl}, and show that its effect is to implement a simple change of frame. This clarifies one of the points raised in~\cite{Bini:2023fiz} when comparing with the post-Newtonian (PN) results. We then provide an explicit PM expression for the waveform in the soft limit, $\omega\to0$, including the first non-universal, $\omega\log\omega$, contribution. Focusing on this regime, we show that the small-velocity limit of our result agrees with the soft limit of the PN waveform of~\cite{Bini:2023fiz}, provided that the two quantities are written in the same asymptotic frame. 
  Performing the BMS supertranslation that, as discussed in~\cite{Veneziano:2022zwh}, is responsible for the $\mathcal O(G)$ static contribution to the asymptotic field employed in the PN literature, we find agreement between the amplitude-based and the PN soft waveform up to and including $G^3/c^5$ order.
}

\begin{document}
	
\maketitle	

\section{Introduction}
\label{sec:intro}

Since 2017~\cite{Damour:2017zjx} there has been a renewed interest in using QFT techniques to study {\em classical} gravitational binaries. The trademark of this approach is to begin the investigation from the scattering problem, \textit{i.e.}~open rather than bound systems, where questions can be directly formulated in terms of the $S$-matrix. In this way, it is possible to use amplitudes as starting points to calculate observables that characterize binary scatterings.\footnote{In addition, for some quantities it is possible to give an explicit map between open and bound orbits~\cite{Kalin:2019rwq,Kalin:2019inp,Cho:2021arx}. Alternatively the scattering results can be used to inform an Effective One Body models~\cite{Buonanno:1998gg,Buonanno:2000ef} as recently done in~\cite{Khalil:2022ylj,Damour:2022ybd,Rettegno:2023ghr}.} Several techniques have been developed to extract from the full quantum amplitudes the classical results of interest~\cite{Neill:2013wsa,Bjerrum-Bohr:2018xdl,Cheung:2018wkq,Kosower:2018adc,KoemansCollado:2019ggb}, including matching the QFT results with an EFT description~\cite{Neill:2013wsa,Cheung:2018wkq}, using the KMOC framework~\cite{Kosower:2018adc} or the eikonal exponentiation~\cite{Amati:1987wq,Amati:1987uf,KoemansCollado:2019ggb,DiVecchia:2023frv}. The QFT based/inspired approach has lead to several explicit new results starting from the conservative deflection angle at 3PM~\cite{Bern:2019nnu} and then considering  more general observables and increasing the precision: focusing on the scattering of Schwarzschild black holes see for instance~\cite{Bern:2019crd,Bern:2021dqo,Bern:2021yeh,Damgaard:2023vnx} for amplitudes and \cite{Kalin:2020mvi,Kalin:2020fhe,Dlapa:2021npj,Kalin:2022hph,Dlapa:2022lmu} for the worldline effective theory. A general feature of the results obtained in this way is that they are valid for generic relative velocities, since Lorentz invariance is kept manifest throughout the calculations. As a non-trivial check, one can expand the PM expressions obtained in the small velocity limit and compare them, in the common regime of validity, with the results available in the PN literature~\cite{Blanchet:2013haa,Bini:2021gat,Bini:2021jmj,Bini:2022enm}. This comparison has been so far relatively straightforward as the physical observables considered in the references above were quantities with a simple dependence on the reference frame, such as the deflection angle, the impulse of the black holes, the total energy and angular momentum carried away by the gravitational radiation.

In this paper we focus on a more complicated scattering observable: the gravitational waveform itself. The leading PM result was obtained within the GR framework long ago \cite{Kovacs:1977uw,Kovacs:1978eu} and, more recently rederived in~\cite{Jakobsen:2021smu,Mougiakakos:2021ckm} by using a worldline theory one-point function equivalent to the classical limit of the tree-level 5-point scattering amplitude~\cite{Goldberger:2016iau,Luna:2017dtq,Mogull:2020sak} with two incoming and two outgoing black holes and an extra graviton in the final state. The comparison with the PN waveform at the leading order for small velocities was already discussed in~\cite{Kovacs:1978eu}, but a systematic comparison with the PN results obtained in the Multipolar-post-Minkowskian (MPM) expansion~\cite{Blanchet:1987wq,Damour:1990gj,Blanchet:1992br,Blanchet:2013haa} has been started only recently~\cite{Bini:2023fiz}. As can be expected, the comparison is trickier because of the dependence on the precise reference frame. The complication becomes fully manifest when including also the first subleading PM order which requires knowing the classical limit of the same five-point amplitude, but now at one-loop. This quantity was obtained in~\cite{Brandhuber:2023hhy,Herderschee:2023fxh,Elkhidir:2023dco,Georgoudis:2023lgf}, where, despite the rather complicated form of the result, it was shown that it has the expected structure. The leading contribution in the amplitude as $\hbar \to 0$ limit is more singular than the tree-level, but it is killed when extracting the classical physical observable following the approaches mentioned above. So in~\cite{Brandhuber:2023hhy,Herderschee:2023fxh,Elkhidir:2023dco,Georgoudis:2023lgf}, the waveform was directly identified with the subleading term of the amplitudes, but it was pointed out in~\cite{Caron-Huot:2023vxl} that this is not the full story because the KMOC approach implies that the waveform includes also an extra contribution associated to cuts of the 5-point amplitude with two intermediate massive particles. In a parallel development~\cite{Bini:2023fiz} initiated a detailed comparison between the PN expansion of the subleading PM result of~\cite{Brandhuber:2023hhy,Herderschee:2023fxh,Elkhidir:2023dco,Georgoudis:2023lgf} and the MPM result for hyperbolic orbit pointing out some mismatches between the two approaches.

It may be natural to think that the two observations above are related and that agreement between the amplitudes and the MPM waveforms would be restored once the contribution pointed out in~\cite{Caron-Huot:2023vxl} is included. We investigated this possibility by focusing on the soft expansion of the full waveform reaching an order where all the points raised in~\cite{Bini:2023fiz} are visible. We find that agreement between the two approaches is indeed restored, but in a subtler way than guessed above. Let us briefly summarize the tensions highlighted in~\cite{Bini:2023fiz}:
\begin{enumerate}
\item The amplitudes (KMOC) and the MPM results are naturally written in two different frames that differ by a rotation of $\Theta/2$, {\em i.e.}~half the scattering angle. While this difference is irrelevant at leading PM order, it should matter for the NLO comparison, and~\cite{Bini:2023fiz} found that even at the Newtonian level the subleading PM term match only if the two results are interpreted to be in the {\em same} frame used in the MPM approach.
\item Various radiation reaction effects are important in the physical waveform. In the PN expansion they start at 1.5PN order with the leading contribution of the tail effect that is already correctly reproduced by the result of~\cite{Brandhuber:2023hhy,Herderschee:2023fxh,Elkhidir:2023dco,Georgoudis:2023lgf}. However, at 2.5PN order ($1/c^5$ corrections with respect to the Newtonian result), Ref.~\cite{Bini:2023fiz} found discrepancies between the PM and the PN results already in the probe limit (the linear terms in $\nu$ for  small mass-ratio $\nu\ll 1$ limit, see~\eqref{eq:nucr} below). This is rather surprising since this part of the NLO PM waveform is completely determined by the well-known tail effect~\cite{Blanchet:1989ki,Blanchet:1995fr}.
  \item The mismatches found in~\cite{Bini:2023fiz} become even more substantial beyond the probe limit when several contributions, due to the non-linearities of gravity, become relevant. 
\end{enumerate}

In order to shed some light on the points above, we start by studying the cut contribution of~\cite{Caron-Huot:2023vxl} and point out the similarity with a mechanism present in a much simpler observable: the 2PM impulse~\cite{Kosower:2018adc,Herrmann:2021tct}. In that case, an analogous cut contribution is important to reproduce the longitudinal parts of the impulse of the process that are present in the frame aligned with the initial spatial momentum of the black holes, while they would disappear in the frame rotated as in point 1 above. Based on the intuition from the eikonal approach, it is natural to expect that a similar pattern holds for the waveform as well and this is further supported by an observation of~\cite{Aoude:2023dui} when studying the waveform to leading order in the soft limit. If this is the case, the cut contribution of~\cite{Caron-Huot:2023vxl} should take a simple form in impact parameter space. We show that this is the case by following an approach~\cite{DiVecchia:2022piu} used also to calculate observables related to the angular momentum from the eikonal operator~\cite{Cristofoli:2021jas,DiVecchia:2023frv}, obtaining a simple expression for this cut in impact-parameter space. We validate this derivation by performing cross checks both in the soft and in the PN limit, and in particular by checking that in the PN limit it agrees with the impact-parameter Fourier transform of the full momentum-space result for the cut obtained in \cite{Georgoudis:2023ozp} (see also~\cite{Bohnenblust:2023qmy} and v4 of~\cite{Brandhuber:2023hhy,Herderschee:2023fxh}). This provides a first-principle derivation for the rotation mentioned in point 1, so we then move on to the analysis of the disagreement between the amplitudes and the MPM results in the PN limit.

Since, so far, it does not seem possible to write a manageable expression for the full PM waveform, we focus our attention on the limit of soft radiation, $\omega\to0$ (or equivalently the early/late time behavior of the waveform). This regime has been under intense investigation since the seminal papers by Weinberg~\cite{Weinberg:1964ew,Weinberg:1965nx} and has more recently been discussed in a series of papers~\cite{Laddha:2018vbn,Sahoo:2018lxl,Saha:2019tub,Sahoo:2021ctw} both from the classical and the amplitudes point of view. The first three terms in the soft expansion, $\frac{1}{\omega}$, $\log\omega$ and $\omega(\log\omega)^2$, are universal in the sense that they can be obtained by acting with an appropriate operator on the elastic result. We extend our analysis to include the next term, $\omega\log\omega$, providing an explicit NLO PM expression for it. Then we reassess points 2 and 3 above by using this result: by following~\cite{Bini:2023fiz}, it is instructive to focus on the difference between the PN results obtained from the soft waveform as derived from amplitudes and those obtained from the soft MPM waveform. This difference does not vanish at order $1/c^5$ for the non-universal $\omega\log\omega$ soft term already at  leading order in the probe limit which, as mentioned above, is entirely determined by the tail effect. However the ${\mathcal O}(\nu)$ mismatch we found is much simpler than the one reported in Eq.~(9.13) of~\cite{Bini:2023fiz}: with our result, one can reabsorb such mismatch with an appropriate choice of the time-origin of the frame (the freedom to perform such shifts was already highlighted in \cite{Bini:2023fiz}). This solves the discrepancy presented in point 2.

We then move to point 3 in the list above by including the ${\mathcal O}(\nu^2)$ terms in our analysis. Again, we find a simpler result than Eq.(9.13) of~\cite{Bini:2023fiz} for the difference between the amplitudes and MPM soft waveforms, however at this order we cannot reabsorb the mismatch simply by a time translation. Note that at this level we start probing in detail both formalisms: on the MPM side various effects become important, including the use of retarded propagators and the non-linearity of gravity (see the review by Blanchet~\cite{Blanchet:2013haa} for a discussion of these points), while on the amplitudes side the corresponding contributions come from ``rescattering''-diagrams where one can isolate a Compton-like subamplitude. In particular on the MPM side one has to take into account some non-linear relation between near- and intermediate-zone multipoles (see for instance~\eqref{Uijklnonlinear}). In evaluating these contribution at the PM order we are interested in, one needs to use the {\em static} contribution to the waveform which depends on the choice of BMS supertranslation frame~\cite{Veneziano:2022zwh}. This issue has played an important role in the calculation of the radiated angular momentum~\cite{Damour:2020tta,Manohar:2022dea, DiVecchia:2022owy,DiVecchia:2022piu,Mao:2023evc,Riva:2023xxm,Compere:2023qoa,Ashtekar:2023zul,Javadinezhad:2023mtp} which depends on the static part of the LO waveform. It is thus natural to expect that the same issue is relevant for the NLO waveform at least for the terms where the static terms are important. The most natural guess is that the NLO waveform obtained from the amplitudes approach is written in the ``canonical'' BMS frame, as defined in~\cite{Veneziano:2022zwh}, while the MPM result is in the frame used in~\cite{Damour:2020tta}. Since from~\cite{Strominger:2014pwa,Veneziano:2022zwh} we know the supertranslation connecting this two frames, it is straightforward to compare the two results in the {\em same} BMS frame: we find that the mismatch mentioned in point~3 above then completely disappears  when using our soft result for the amplitude-based waveform! This shows that the disagreement arises due to two different choices of asymptotic frames, which differ by a BMS supertranslation, rather then a Poincar\'e transformation as for the first two points.

The paper is structured as follows. In Section~\ref{sec:elastic}, as preparatory work, we recall some results about the elastic process in particular to present a derivation of the 2PM longitudinal impulse which is instructive for our purposes. In Section~\ref{sec:PMwaveform} we review the results on the 1-loop 5-point amplitude~\cite{Brandhuber:2023hhy,Herderschee:2023fxh,Elkhidir:2023dco,Georgoudis:2023lgf} and the cut contribution highlighted in~\cite{Caron-Huot:2023vxl}. We obtain an explicit expression for the latter in term of known leading PM quantities. The key point of the derivation is to move to impact parameter space before doing the integrals over the cut propagators, which allows to simplify the expression since only the long-range contributions survive. In Section~\ref{sec:Soft} we perform the soft limit of the NLO PM waveform derived from the amplitudes approach including the first non-universal term. In Section~\ref{sec:compPN} we compare our soft results in the small velocity limit with those obtained in the PN MPM approach and reassess the mismatch with the amplitude-based waveform pointed out by \cite{Bini:2023fiz}. We provide a simple symmetry argument showing that the 2.5PN contribution to each multipole does not receive corrections from the cut highlighted in~\cite{Caron-Huot:2023vxl}. However, we show that the results in the two approaches match after making sure that they are written in the same BMS frame.
We conclude in Section~\ref{sec:outlook} by summarizing the main results and discussing what in our opinion are the most interesting possible future developments.

\paragraph*{Note added.}
At the final stages of this work, we became aware of the parallel work of the collaboration of Donato Bini, Thibault Damour, Stefano De Angelis, Andrea Geralico, Aidan Herderschee, Radu Roiban and Fei Teng who independently derived the full PM soft limit of the NLO waveform, including the first non-universal term, $\omega\log\omega$, and find agreement with our results \cite{toapFei}. They also derived the mismatch between the MPM and the amplitudes based results at 2.5PN order, reproducing exactly Eq.~\eqref{eq:mismst} below.

\section{From Amplitudes to Classical GR Observables: the Elastic Case}
\label{sec:elastic}

We focus on minimally coupled massive scalars with a ``large'' (\textit{i.e.}~classical) mass so they can be used to model Schwarzschild black holes. In this setup, standard perturbation theory breaks down and Feynman diagrams involving only few gravitons do not provide a faithful description of the scattering between two black holes, see for instance~\cite{DiVecchia:2023frv} for a discussion in the framework of the eikonal approach. 
However, it is possible to rearrange the standard QFT perturbative expansion so as to extract from each order in the Newton constant\footnote{We shall also use the symbol $\kappa = \sqrt{8\pi G}$.} $G$ the contributions relevant to the classical Post-Minkowskian (PM) expansion and to make contact with GR observables. This step can be done in a variety of ways: by matching with an effective field theory~\cite{Goldberger:2004jt,Porto:2005ac,Porto:2016pyg,Cheung:2018wkq,Cristofoli:2020uzm}, in the KMOC approach~\cite{Kosower:2018adc,Herrmann:2021lqe,Herrmann:2021tct,Damgaard:2023vnx,Damgaard:2023ttc} and by using the eikonal exponentiation (see~\cite{DiVecchia:2023frv} and references therein) among others. Focusing on the last two approaches, the idea is to define the incoming state by considering two well-separated objects described by a wave-packet superposition of plane-wave states,
\begin{equation}\label{}
	| \text{in} \rangle = |1\rangle \otimes |2\rangle\,,
\end{equation}
with
\begin{equation}\label{eq:in-p}
\begin{aligned}
	|1\rangle &=\int\! 2\pi\delta(p_1^2+m_1^2) \theta(-p_1^0) \frac{d^Dp_1}{(2\pi)^D}
	\,\varphi_1(-p_1) \,e^{ib_1\cdot p_1}|\!-\!p_1\rangle \equiv \int_{-p_1} \varphi_1(-p_1) \,e^{ib_1\cdot p_1}|\!-\!p_1\rangle \\
	|2\rangle &=
	\int\! 2\pi\delta(p_2^2+m_2^2) \theta(-p_2^0) \frac{d^Dp_2}{(2\pi)^D}\,\varphi_2(-p_2)\,e^{ib_2\cdot p_2}|\!-\!p_2\rangle \equiv \int_{-p_2} \varphi_2(-p_2)\,e^{ib_2\cdot p_2}|\!-\!p_2\rangle \,.
      \end{aligned}
\end{equation}

The classical value of an observable ${\cal O}$ is obtained by calculating the leading order contribution in the classical limit to the expectation value
\begin{equation}
  \label{eq:cl-ev}
  \langle {\cal O} \rangle \equiv \langle \text{out}| {\cal O} |\text{out}\rangle\,,\qquad  |\text{out}\rangle = S |\text{in}\rangle\,.
\end{equation}
It is convenient to formally write the final state in the plane wave basis and in the elastic case we have
\begin{equation}
  \label{eq:outin}
  |\text{out}\rangle = \int_{-p_1} \int_{-p_2} \int_{p_3} \int_{p_4} \varphi_1(-p_1)\varphi_2(-p_2)\,e^{i(b_1\cdot p_1+b_2\cdot p_2)} |p_3,p_4\rangle \langle p_3,p_4| S |-p_1,-p_2\rangle + \cdots 
\end{equation}
and one can evaluate the $S$-matrix with standard QFT methods. The KMOC approach  exploits the fact that, for classical observables,\footnote{When considers quantum observables, such as the average number of gravitons  
emitted during a scattering process, one would find results that diverge in the classical limit. Here, we will focus on classical observables, such as the deflection angle and the waveform.} the expectation value~\eqref{eq:cl-ev} has finite classical limit order by order in the $G$-expansion. In order to make this manifest, it is convenient to introduce the following scaling for the momenta
\begin{equation}\label{eq:lali}
	p_1 \sim p_2 \sim \mathcal O(\lambda^0)\,,\qquad
	q_i \sim k \sim \mathcal O (\lambda)\,,\quad \mbox{as }\quad \lambda\to 0\;,
\end{equation}
where $\lambda$ is proportional to $\hbar$, and so it is small, $p_i$ are the external momenta of the black holes as in~\eqref{eq:in-p}, while with $q_i$ and $k$ we will indicate the momenta of virtual and real gravitons respectively. For later convenience we introduce the notation below to indicate the behavior of an arbitrary quantity $\mathcal Q$ under the scaling~\eqref{eq:lali}
\begin{equation}\label{}
	\mathcal Q^{[-j]} \sim \mathcal O(\lambda^{-j})\,.
\end{equation}
When computing~\eqref{eq:cl-ev} in the KMOC approach, we will need to extract at each order in $G$ the first non-trivial contribution to the expectation value in the limit~\eqref{eq:lali} and to rewrite it in terms of classical quantities (which can be done since all factors of $\lambda$ or of $\hbar$ should have canceled). Instead, in the eikonal approach one uses a classical approximation for the $S$-matrix in order to obtain an expression for the out-states and for the expectation values, see~\eqref{eq:cl-ev}. This classical $S$-matrix arises by exponentiating the divergent terms as $\hbar\to 0$, which is done by working in impact parameter space. While at the LO in the PM expansion the results obtained in the two approaches can be compared directly, it turns out that at subleading order the eikonal impact parameter $b_e$ is different from the one used in the KMOC approach which we indicate as $b_J$, to stress that it is directly related to the initial angular momentum $J$ of the binary in the center of mass frame. In order to spell out this point, which will play an important role in the analysis of the waveform, let us first introduce our conventions on the kinematics (following closely~\cite{DiVecchia:2023frv}) starting first from the elastic scattering and then including radiation.

For the elastic tree-level $2\to2$ amplitude we have, 
\begin{equation}\label{A04pt}
	\mathcal A_0^{(4)}(p_1,p_2;q)
	=
	\begin{gathered}
		\begin{tikzpicture}
			\path [draw, ultra thick, blue] (-4,2)--(-2,2);
			\draw[<-] (-4,2.12)--(-3.7,2.12);
			\draw[<-] (-2,2.12)--(-2.3,2.12);
			\path [draw, ultra thick, green!60!black] (-4,1)--(-2,1);
			\draw[<-] (-4,.88)--(-3.7,.88);
			\draw[<-] (-2,.88)--(-2.3,.88);
			\filldraw[black!20!white, thick] (-3,1.5) ellipse (.5 and .8);
			\draw[thick] (-3,1.5) ellipse (.5 and .8);
			\draw[->] (-2.3,1.3)--(-2.3,1.7);
			\node at (-2.3,1.5)[right]{$q$};
			\node at (-4,2)[left]{$p_1$};
			\node at (-4,1)[left]{$p_2$};
			\node at (-2,1)[right]{$p_3$};
			\node at (-2,2)[right]{$p_4$};
		\end{tikzpicture}
	\end{gathered}
	=
	\mathcal A_0^{(4)[-2]}
	(\bar p_1,\bar p_2;q)+\mathcal O(\lambda^{0})\,,
\end{equation}
with
\begin{equation}\label{eq:pbarp}
	q^\mu=p^\mu_1+p^\mu_4=-p^\mu_2-p^\mu_3\,,\quad
	\bar p^\mu_1=\frac{p^\mu_4-p^\mu_1}{2} = -p^\mu_1 +\frac{q^\mu}{2}\,,\quad
	\bar p^\mu_2=\frac{p^\mu_3-p^\mu_2}{2} = -p^\mu_2 - \frac{q^\mu}{2}\,,
\end{equation}
where $q\sim \hbar/b_e \sim \hbar/b_J$ is quantum and so it is also the difference between $\bar{p}_{1,2}$ and $-p_{1,2}$. 

\subsection{PM impulse}
\label{ssec:impulse}

Let us now see how the classical impulse $Q^\mu$ describing the total momentum exchanged in the process emerges in the KMOC approach. In order to do this we need to calculate the expectation value~\eqref{eq:cl-ev}, for the observable representing the variation of the momentum of one of the massive states ({\rm i.e.} the impulse). For instance, denoting by $P_1^\mu$ the operator measuring the momentum of particle 1, we have
\begin{equation}
  \label{eq:impul}
  Q^\mu = \langle \text{out} | P_1^\mu | \text{out} \rangle - \langle \text{in} | P_1^\mu | \text{in} \rangle= Q_{\text{1PM}}^\mu +  Q_{\text{2PM}}^\mu + {\cal O}(G^3)\,.
\end{equation}
In order to write the results of the expectation values in terms of classical quantities we need to perform a Fourier transform to impact parameter space
\begin{equation}\label{eq:ft}
  \operatorname{FT}[{\cal O}] 
  = \int \frac{d^D q}{(2\pi)^D} \, 2\pi \delta(2p_1\cdot q)\,2\pi\delta(2p_2\cdot q)\, {\cal O}(q)\,e^{ib \cdot q}\,,
\end{equation}
where ${\cal O}$ is a generic observable. When the Fourier transform \eqref{eq:ft} is performed on a quantity whose leading behavior is classical, we can neglect the difference between $p_i$ and $\bar{p}_i$ and the shifts $\pm q^2$ that would appear in the delta functions when writing the full on-shell conditions for the final states. These differences have an extra factor of $\hbar$ and so they cannot change the leading classical term. However the interpretation of the impact parameter $b$ appearing in~\eqref{eq:ft} depends on how the classical quantity has been extracted from the  amplitude, which in general contains terms which are more divergent than the classical ones as $\hbar \to 0$. When applying~\eqref{eq:ft} to observables obtained in the KMOC approach, the impact parameter should be interpreted as $b_J$, {\rm i.e.} the quantity directly related to the orbital angular momentum $J = |\vec{p}\,| b_J$. We review this point below in the simple example of the 2PM impulse~\cite{Herrmann:2021tct}. Let us stress that $b_J$ is the impact parameter appearing also at subleading PM orders whenever the subtraction leading to the classical result is done by using the full unitarity cuts~\cite{Herrmann:2021tct,Damgaard:2021ipf,Brandhuber:2021eyq}.

Coming back to the impulse in the KMOC approach, we can compute perturbatively~\eqref{eq:impul} by expanding $S=1+i T$ and writing the matrix elements of $T$ in terms of amplitudes. 
Then, at the first order in $G$, the expectation value gets contributions only from terms linear in $T$ and so it can be written in terms of a tree-level amplitude, which we can approximate at the leading order in the regime~\eqref{eq:lali}. The operator $P_1^\mu$ inserted in the expectation value yields either $p_1$ or $p_4$ for the terms linear in $T$ and so we get
\begin{subequations}    \label{eq:2.9}
  \begin{equation}
    \label{eq:impelkm}
      Q_\text{1PM}^\mu= i\operatorname{FT}\left[(p_1+p_4)^\mu  \mathcal A_0^{(4)[-2]}\right] = i\operatorname{FT}\left[q^\mu \mathcal A_0^{(4)[-2]}\right],
  \end{equation}
where, for minimally coupled massive scalars,
 \begin{equation}\label{4ptGR}
	\mathcal A_0^{(4)[-2]}
	(p_1,p_2;q)
	=
	\frac{32 \pi  G {m}_1^2 {m}_2^2 \left(\sigma^2 -\frac{1}{D-2}\right)}{q^2}\,,
\end{equation}
\end{subequations}
with
\begin{equation}
  \label{eq:sigE}
  \sigma= -\frac{p_1 \cdot p_2}{m_1 m_2}\,, \qquad E=\sqrt{-(p_1+p_2)^2} = \sqrt{m_1^2+2 m_1 m_2 \sigma + m_2^2}\,.
\end{equation} 
By using~\eqref{eq:2.9} we have
\begin{equation}\label{Q1PMasintegral}
 Q_{\text{1PM}}^\mu = 
  - \frac{b_J^\mu}{b_J} \, \frac{4G m_1 m_2 \left(\sigma^2-\frac{1}{2}\right)}{b_J\sqrt{\sigma^2-1}}\,.
\end{equation}
Notice that in order to restore length and mass units in the equation above, one does not need factors of $\hbar$: as advertised, expectation values of classical observables are independent of it. Thus we can interpret \eqref{Q1PMasintegral} as the classical result at order $G$, \textit{i.e.}~at leading order in the PM expansion.

The contributions coming from elastic diagrams involving more exchanged gravitons will yield subleading PM corrections. Let us discuss the 2PM correction in some detail as in the next section we will employ a similar approach to study the subleading contribution to the waveform. The second term in the PM expansion of the impulse, $Q_\text{2PM}^\mu$,
contains two contributions~\cite{Kosower:2018adc,Herrmann:2021tct}.\footnote{A similar analysis has recently been discussed for the scattering of spinning particles~\cite{Gatica:2023iws}} One comes from the terms linear in $T$ in the expectation values~\eqref{eq:impul} and is proportional to a scattering amplitude which now has to be evaluated at the second order in $G$, while the second one comes from the term quadratic in $T$ where each factor is evaluated at order $G$, 
\begin{equation}\label{1loopkerelQ}
  Q_{\text{2PM}}^\mu = i\operatorname{FT}\left[p_4  \mathcal{A}_1^{(4)} + p_1  \mathcal{A}_1^{(4)\ast} - i \int (\ell - p_1)
  \mathcal{A}_0^{(4)}(p_1,p_2;\ell) \mathcal{A}_0^{(4) \ast}(p_3,p_4;q-\ell) dL(\ell) \right]\,,
\end{equation}
where $\mathcal{A}_1^{(4)}$ ($\mathcal{A}_0^{(4)}$) is the elastic 4-point 1-loop (tree-level) amplitude and $\int dL(\ell)$ indicates an integral over $\ell$ where the cut propagators are on-shell
\begin{equation}
  \label{eq:dLell}
  \int dL(\ell) = \int \frac{d^D\ell}{(2\pi)^D}\, 2\pi\delta(2p_1\cdot \ell-\ell^2)\, 2\pi\delta(2p_2\cdot \ell+\ell^2) \;.
\end{equation}
It is convenient to rewrite this result in terms of the real and the imaginary part of $\mathcal{A}_1^{(4)}$. 
Thanks to unitarity $-i(T-T^\dagger) = T^\dagger T$, the imaginary part can be combined with the last term in~\eqref{1loopkerelQ} yielding
\begin{equation}\label{QResmu}
	Q_\text{2PM}^\mu =  i\operatorname{FT}\left[q^\mu  \operatorname{Re}\mathcal A_1^{(4)[-1]} + i s^\mu\right],
\end{equation}
where
\begin{gather}
	\label{eq:Q2im}
	s^\mu = \int \left(\frac{p^\mu_4-p^\mu_1}{2} - (\ell^\mu - p^\mu_1) \right) \mathcal{A}_0^{(4)}(p_1,p_2;\ell) \mathcal{A}_0^{(4)\ast}(p_3,p_4;q-\ell) dL(\ell)\,.
\end{gather}
The contribution from $\operatorname{Re}\mathcal{A}^{(4)}_1$ in \eqref{QResmu} follows the same pattern we saw at tree level: its leading behavior in the regime~\eqref{eq:lali} (which is ${\mathcal O}(\lambda^{-1})$) ensures that the final contribution to the impulse is classical. Focusing again on GR for instance, we have
\begin{equation}
  \label{eq:Q2re}
  \operatorname{Re}\mathcal{A}^{(4)[-1]}_1 = 2 \pi G m_1^2 m_2^2 (m_1+m_2) \frac{3 \pi (5\sigma^2 -1)}{q}\,,
\end{equation}
and thus
\begin{equation}\label{eq:Q2fre}
    i\operatorname{FT}\left[q^\mu  \operatorname{Re}\mathcal A_1^{(4)[-1]} \right] = -\frac{b^\mu_J}{b_J} \frac{3\pi G^2 m_1 m_2 (m_1+m_2) (5\sigma^2-1)}{4 b_J^2 \sqrt{\sigma^2-1}} = -\frac{b^\mu_J}{b_J} \,   Q_{\text{2PM}} \,.
\end{equation}
As anticipated, the contribution above is classical since it has the dimension of a mass without the need of introducing $\hbar$. Instead, the term involving $s^\mu$ in \eqref{QResmu} naively seems to have contributions that scale as ${\mathcal O}(\lambda^{-2})$ by \eqref{eq:Q2im},
where the two terms in the round parenthesis are separately $\mathcal{O}(\lambda^0)$. However, the leading term cancels and we obtain
\begin{equation}
    \label{eq:Q2im2}
    \tilde{s}^\mu = \operatorname{FT} \left[\int \left(\tfrac 12 \, q^\mu-\ell^\mu\right) \mathcal{A}_0^{(4)}(p_1,p_2;\ell) \mathcal{A}_0^{(4)\ast}(p_3,p_4;q-\ell)
     dL(\ell)\right]\;.
\end{equation}

Eq.~\eqref{eq:Q2im2} was evaluated explicitly in~\cite{Herrmann:2021tct} by doing the integral over $\ell$ first and then performing the Fourier transform. It is instructive to reproduce the same result by doing the two operations in the opposite order as this approach will be useful in the next section when discussing a similar cut contribution in the inelastic case. By using the variables $\bar{p}_i$~\eqref{eq:pbarp}, so as to exploit the simpler classical expansion in~\eqref{A04pt}, we have
\begin{equation}
  \label{eq:Q2im3}
\begin{aligned}
  \tilde{s}^\mu= \operatorname{FT} &\Big[\int\frac{d^D\ell}{(2\pi)^D}\,
    \left(\tfrac 12 q^\mu-\ell^\mu \right)
      \mathcal A_0^{(4)}(p_1,p_2;\ell)\,\mathcal A_0^{(4)\ast}(p_3,p_4;q-\ell)
       \\  & \times 2\pi\delta(2\bar p_1\cdot \ell+\ell\cdot(\ell-q))2\pi\delta(2\bar p_2\cdot \ell-\ell\cdot(\ell-q)) \Big].
     \end{aligned}
\end{equation}   
At the first order in the classical limit~\eqref{eq:lali} we can use~\eqref{A04pt} and approximate the delta functions by keeping only the linear terms in $\bar{p}_i$. However this contribution vanishes \cite{Kosower:2018adc,Herrmann:2021tct}, by using the delta functions and the fact that the integrand becomes odd in $\ell\to q-\ell$ at this order. Thus the first non-trivial term follows from the expansion of the delta functions
\begin{equation}
  \label{eq:expdel1}
\begin{aligned}
  \delta(2\bar p_1\cdot \ell+\ell\cdot(\ell-q)) & \delta(2\bar p_2\cdot \ell-\ell\cdot(\ell-q)) = \delta(2\bar p_1\cdot \ell) \delta(2\bar p_2\cdot \ell) \\ & + \ell \cdot (\ell-q) \left(\delta'(2\bar p_1\cdot \ell) \delta(2\bar p_2\cdot \ell)-\delta(2\bar p_1\cdot \ell) \delta'(2\bar p_2\cdot \ell) \right)+\cdots\,,
\end{aligned}
\end{equation}
while, thanks to~\eqref{A04pt}, there are no NLO contribution from the tree amplitude. Since the result is classical, we may now approximate $\bar p_i^\mu \simeq -p_i^\mu$ ($i=1,2$), 
and by using the fact that $\mathcal A_0^{(4)[-2]}$ is real write
\begin{equation}
  \label{eq:Q2im4}
\begin{aligned}
  \tilde{s}^\mu = \operatorname{FT} & \Big[\int\frac{d^D\ell}{(2\pi)^D}\,
    \left(\ell^\mu-\tfrac 12 q^\mu \right)
      \mathcal A_0^{(4)[-2]}(\ell)\,\mathcal A_0^{(4)[-2]}(q-\ell)
       \\  & \times (2\pi)^2 \ell (\ell-q) \left(\delta'(2  p_1 \cdot \ell) \delta(2 p_2\cdot \ell)-\delta(2  p_1\cdot \ell) \delta'(2  p_2\cdot \ell) \right) \Big]
     \end{aligned}
\end{equation}   
up to quantum corrections.
Let us introduce the following variables to indicate the velocities of the massive particles
\begin{equation}
  \label{eq:vel}
    p^\mu_1= - m_1 v^\mu_1\,,\qquad
     p^\mu_2= - m_2 v^\mu_2 \,, \qquad
     \sigma = -v_1\cdot v_2\,
  \end{equation}
and the dual velocities
\begin{equation}\label{eq:dualvel}
	\check v^\mu_1 = \frac{\sigma v^\mu_2-v^\mu_1}{\sigma^2-1}\,,\qquad
	\check v^\mu_2 = \frac{\sigma v^\mu_1-v^\mu_2}{\sigma^2-1}\,,\qquad
\end{equation}
which satisfy $\check v_i \cdot v_j = -\delta_{ij}$, so that
\begin{equation}
  \label{eq:delpr}
\delta'(2 p_1\cdot \ell) = \frac{\check v^\rho_1}{2 m_1} \frac{\partial}{\partial \ell^\rho} \delta(2 p_1\cdot \ell)\,, \qquad
\delta'(2 p_2\cdot \ell) = \frac{\check v^\rho_2}{2 m_2} \frac{\partial}{\partial \ell^\rho} \delta(2 p_2\cdot \ell)\,.
\end{equation}
Using the above expressions to rewrite the delta functions appearing in~\eqref{eq:Q2im4}, and integrating by parts, we can check that the only term giving a non-trivial ${\cal O}(\lambda^0)$ contribution is obtained when the derivative acts on $\ell^\mu$. Thus we get
\begin{equation}\label{}
  s^\mu =
	\left(
	\frac{\check v_1^\mu}{2 m_1}- \frac{\check v_2^\mu}{2 m_2}
	\right)
	\int\frac{d^D\ell}{(2\pi)^D}\,
	\mathcal A_0^{(4)[-2]}(\ell)\,\ell\cdot(q-\ell)\mathcal A_0^{(4)[-2]}(q-\ell)\,2\pi\delta(2\bar p_1\cdot \ell)2\pi\delta(2\bar p_2\cdot \ell)
\end{equation}
 Going to impact-parameter space, the convolution factorizes (see e.g.~\cite{DiVecchia:2023frv} for more details),
and one is left with
\begin{equation}\label{ImaginaryPartFT}
  \tilde{s}^\mu =-
	\left(
	\frac{\check v_1^\mu}{2 m_1}- \frac{\check v_2^\mu}{2 m_2}
	\right)
	Q_\text{1PM}^2\,.
\end{equation}
Inserting \eqref{eq:Q2fre} and \eqref{ImaginaryPartFT} in \eqref{QResmu}, we finally have
\begin{equation}\label{Q2PMvector}
	Q^\mu_{\text{2PM}}
	= -\frac{b_J^\mu}{b_J}\,Q_{\text{2PM}} + \left(
	\frac{\check v_1^\mu}{2 m_1}- \frac{\check v_2^\mu}{2 m_2}
	\right) Q^2_\text{1PM}\,.
\end{equation}

The longitudinal terms in \eqref{Q2PMvector} are thus induced by the following transformation of the basis vectors,
\begin{equation}
	\label{barp1barp2}
		m_1 v_1^\mu \mapsto \tilde m_1 \tilde{u}_1^\mu = \tilde p_1^\mu = -p_1^\mu + \frac{1}{2}\,Q^\mu\,,\qquad
		m_2 v_2^\mu \mapsto \tilde m_2 \tilde{u}_2^\mu = \tilde p_2^\mu = -p_2^\mu - \frac{1}{2}\,Q^\mu\,,
\end{equation}
and
\begin{equation}\label{betransf}
	b_J^\mu \mapsto b_e^\mu = b_J^\mu - \left(
	\frac{\check v_1^\mu}{2 m_1}-\frac{\check v_2^\mu}{2m_2}
	\right)
	Q\, b_J\,,
\end{equation}
where $\tilde p_{1,2}$ and $b_e^\mu$ are defined in such a way that, while $p_{1,2}\cdot b_J=0$,
\begin{equation}\label{}
	\tilde p_{1,2}\cdot b_e = 0\,. 
\end{equation}
Notice that the momenta $\tilde{p}_i$ and $-p_i$ differ by a classical quantity in contrast to what happens with the variables $\bar p_i$ defined in~\eqref{eq:pbarp}. In analogy to what was done in~\eqref{eq:vel} and~\eqref{eq:dualvel}, we introduce for later convenient also the velocities $u_i$
\begin{equation}
  \label{eq:uel}
    \bar p^\mu_1= - \bar m_1 u^\mu_1\,,\qquad
     \bar p^\mu_2= - \bar m_2 u^\mu_2 \,, \qquad
     y = -u_1\cdot u_2\,
  \end{equation}
and the dual velocities
\begin{equation}\label{eq:dualuel}
	\check u^\mu_1 = \frac{y u^\mu_2-u^\mu_1}{y^2-1}\,,\qquad
	\check u^\mu_2 = \frac{y u^\mu_1-u^\mu_2}{y^2-1}\,,\qquad
\end{equation}
which satisfy $\check u_i \cdot u_j = -\delta_{ij}$. Coming back to the impulse the difference between the norms $b_e$ and $b_J$ is of order $G^2$ (and similarly between $m_i$ and $\tilde{m}_i$). Thus, at 2PM order one can regard the transformation of $b_J^\mu$ as a rotation and by using~\eqref{betransf} in~\eqref{Q2PMvector} we have
\begin{equation}\label{QuptoorderG2}
Q^\mu 
=
- \frac{b_e^\mu}{b_e}\left(
Q_\text{1PM} + Q_\text{2PM}
\right)
+
\mathcal O(G^3)\,.
\end{equation}
The infinitesimal action of \eqref{barp1barp2}, \eqref{betransf} can be written as follows,
\begin{equation}\label{transformationDEFQ}
	\bar\delta Q^{\mu}
	= Q_\text{1PM}\, \bar\partial Q^\mu  + \mathcal O(G^3)\,,
\end{equation}
where
\begin{equation}\label{myDiffOp}
	\bar \partial = - \frac{b_J^\alpha}{b_J}\left(
	\frac{1}{2 m_1}
	\frac{\partial}{\partial v_1^\alpha}
	-
	\frac{1}{2 m_2}
	\frac{\partial}{\partial v_2^\alpha}
	\right)
	-
	b_J
	\left(
	\frac{\check v_1^\alpha}{2 m_1}
	-
	\frac{\check v_2^\alpha}{2 m_2}
	\right)
	\frac{\partial}{\partial b_J^\alpha}\,,
\end{equation}
is a differential operator that preserves the constraints $v_{1,2}\cdot b_J=0$. So we can write the longitudinal contribution as follows
\begin{equation}\label{s-change}
  \tilde{s}^\mu =
	\bar\delta Q^\mu = Q_\text{1PM} \bar\partial Q^\mu\,.
\end{equation}

It is instructive to explicitly check that acting with $\bar\partial$ as in \eqref{transformationDEFQ} indeed reproduces the longitudinal terms in \eqref{ImaginaryPartFT} both when using the explicit 1PM expression~\eqref{Q1PMasintegral}, where only the $\partial/\partial b_J^\alpha$ part of $\bar\partial$ acts non-trivially, and when using the integral expression \eqref{eq:impelkm}, where instead only the $\partial/\partial v_1^\alpha$, $\partial/\partial v_2^\alpha$ terms act non-trivially.
This interplay is due to the fact that only the sum of the two terms in \eqref{myDiffOp} is a well-defined differential operator on the constrained surface $v_{1,2}\cdot b_J=0$, whereas the two terms are separately ambiguous. 

Let us conclude this section with a brief comparison between the KMOC and the eikonal approach. The basic idea of the latter is to first obtain a classical approximation for the out-state by resumming the leading contributions in the classical limit into an exponential form. In the elastic approximation, we expect
\begin{equation}
  \label{eq:eikexpel}
  |\text{out}\rangle = \left[1 + i \Delta(E,b_e) \right] e^{2i\delta(E,b_e)} |\text{in}\rangle\,,
\end{equation}
where $\Delta$ captures quantum effects while the classical part, scaling as $1/\hbar$, exponentiates. The quantities appearing on the r.h.s. can be determined by performing a formal expansion in $G$ on the $S$-matrix elements $\langle \text{in} |\text{out}\rangle$ and matching the result with perturbative calculations, see~\cite{DiVecchia:2023frv} and references therein. This allows to determine the eikonal phase $\delta(E,b_e)$ in a PM expansion and, for instance, in the case of GR we have at leading and subleading order
\begin{equation}
  \label{eq:delta0GR}
      2\delta_0 = \tilde{\mathcal{A}}_0^{(2)[-2]} = \frac{2 G m_1 m_2 (\sigma^2-\frac{1}{D-2})}{\sqrt{\sigma^2-1}} \frac{\Gamma(-\epsilon)}{(\pi b_e^2)^{-\epsilon}}\,,\qquad
      2\delta_1 = \frac{3\pi G^2m_1m_2(m_1+m_2)(5\sigma^2-1)}{4b_e\sqrt{\sigma^2-1}}\,.
\end{equation}
Let us stress that even the leading contribution captures (after exponentiation) an infinite number of ladder and cross-ladder diagrams,
\begin{equation}\label{}
	1+i \operatorname{FT}
	\begin{gathered}
		\begin{tikzpicture}
			\path [draw, ultra thick, blue] (-4,2)--(-2,2);
			\path [draw, ultra thick, green!60!black] (-4,1)--(-2,1);
			\filldraw[white, thick] (-3,1.5) ellipse (.5 and .8);
			\filldraw[pattern=north east lines, thick] (-3,1.5) ellipse (.5 and .8);
			\draw[thick] (-3,1.5) ellipse (.5 and .8);
		\end{tikzpicture}
	\end{gathered}
	= e^{2i\delta_0}\,.
\end{equation}

Since the matching in~\eqref{eq:eikexpel} is performed on a result with terms that are more divergent than the classical ones, one has to be careful about subleading quantum corrections. The subtraction implied by this procedure requires to interpret the impact parameter appearing in the equation in a different way from the one used in the KMOC approach. A first signal of this can be seen by rewriting the eikonal $S$-matrix in momentum space: the Fourier transform can be performed via a stationary phase argument (and similarly for the transformation from the energy to the time domain) leading to the the impulse and (IR-divergent) Shapiro time delay
\begin{equation}\label{TGR0}
	Q^\mu = \frac{\partial 2\delta}{\partial (b_e)_\mu} \,,\qquad
	T = \frac{\partial 2\delta}{\partial E} = \frac{E}{m_1 m_2} \frac{\partial 2\delta}{\partial \sigma}\,.
\end{equation}
Because of the kinematic constraint $Q=2|\vec{p}\,| \sin(\Theta/2)$ between the impulse and the elastic scattering angle $\Theta$, there is in general a non-linear relation between the angle and $\partial_{b_e}(2\delta)$ contrary to what happens when one takes the derivative of the radial action with respect to $b_J$. By using~\eqref{eq:delta0GR} and~\eqref{TGR0} one then directly obtains \eqref{QuptoorderG2} for GR, with $Q^\mu$ written in terms of $b_e^\mu$ \cite{DiVecchia:2023frv}, and, to leading order,
\begin{equation}\label{TGR}
	T_{\rm 1PM} = 2 G E \,\frac{\sigma \left(\sigma^2-\frac{3-4\epsilon}{2-2\epsilon}\right)}{(\sigma^2-1)^{3/2}} \frac{\Gamma(-\epsilon)}{(\pi b_e^2)^{-\epsilon}}\,.
\end{equation}

\section{From Amplitudes to the PM Waveform}
\label{sec:PMwaveform}
Let us now proceed to study another classical observable, the expectation value of the canonically quantized graviton field in the final state: 
\begin{equation}\label{hmunubasic}
	\langle H_{\mu\nu}(x) \rangle = \langle \text{out}|\int_k \left[
	e^{ik\cdot x} a_{\mu\nu}(k) 
	+
	e^{-ik\cdot x} a^\dagger_{\mu\nu}(k) \right]|\text{out}\rangle\,,
 \end{equation}
 where, as before, the out-state is dictated by the $S$-matrix as in \eqref{eq:cl-ev}. However, compared to \eqref{eq:outin} we now need to also include the first inelastic contribution with one graviton of momentum $k^\mu$ in the final state,
 \begin{equation}
   \label{eq:outin2}
   \begin{aligned}
     &|\text{out}\rangle = \int_{-p_1} \int_{-p_2} \int_{p_3} \int_{p_4} \varphi_1(-p_1)\varphi_2(-p_2) e^{i(b_1\cdot p_1+b_2\cdot p_2)} \\
     &\times \Big [|p_3,p_4\rangle \langle p_3,p_4| S |-p_1,-p_2\rangle + \int_k |p_3,p_4,k\rangle \langle p_3,p_4,k| S |-p_1,-p_2\rangle + \cdots \Big] \,.
   \end{aligned}
\end{equation}
The metric fluctuation sourced by the scattering,
\begin{equation}\label{}
	g_{\mu\nu}(x)-\eta_{\mu\nu} = 2 \sqrt{8\pi G}\, \langle H_{\mu\nu}(x) \rangle= h_{\mu\nu}(x)\,,
\end{equation}
is determined by the so-called wave-shape $W$~\cite{Kosower:2018adc,Cristofoli:2021vyo,Cristofoli:2021jas},
\begin{equation}\label{eq:Wsh}
	\langle\text{out}|a_{\mu\nu}(k)|\text{out}\rangle
	= i \operatorname{FT}[ W_{\mu\nu}(k)]  \equiv i \tilde{W}_{\mu\nu}\,,
      \end{equation}
where the generalization of the Fourier transform \eqref{eq:ft}  reads
\begin{equation}\label{FT5}
	\operatorname{FT}[f] = \!\int \frac{d^Dq_1}{(2\pi)^D}\,\frac{d^Dq_2}{(2\pi)^D}\,(2\pi)^D\delta^{(D)}(q_1+q_2+k)\,2\pi\delta(2p_1\cdot q_1)\,2\pi\delta(2p_2\cdot q_2)\,e^{ib_1\cdot q_1+i b_2\cdot q_2}\,f(q_1,q_2,k)\,.
\end{equation}
In the rest of the paper, we shall employ the following notation for the $5$-point kinematics. We denote
\begin{equation}
	\label{eq:o1o2}
	\omega_1 = -v_1\cdot k\,,
	\qquad 
	\omega_2 =-v_2\cdot k\,,
\end{equation}
so that $\omega_i$ with $i=1,2$ are the frequencies of the emitted graviton in the rest frame of the first or the second massive particle,
and
\begin{equation}\label{eq:kc5}
	q^\mu_1= p^\mu_1+p^\mu_4\,,
	\qquad  q^\mu_2= p^\mu_2+p^\mu_3
	\,,
	\qquad \, q^\mu_1+q^\mu_2+k^\mu=0\,,
\end{equation}
so that
\begin{equation}\label{pbar5pt}
	\bar p_1^\mu = -p_1^\mu + \frac{1}{2}\,q_1^\mu\,,
	\qquad 
	\bar p_2^\mu = -p_2^\mu + \frac{1}{2}\,q_2^\mu
\end{equation}
obey $\bar p_1\cdot q_1=0$, $\bar p_2\cdot q_2=0$. Note that \eqref{pbar5pt} reduces to \eqref{eq:pbarp} in the elastic case $k^\mu=0$, $q_1^\mu= q^\mu=-q_2^\mu$.

Let us also recall the following useful identities
\begin{equation}\label{}
	\quad m_1 \omega_1+m_2\omega_2 = E \omega\,,\qquad
	E |\vec p\,| = m_1 m_2 \sqrt{\sigma^2-1}\,,
\end{equation}
where $E$, $\vec p$ and $\omega$ are, respectively, the total energy, the massive particle's spatial momentum and the graviton's frequency in the center of mass frame.
$\tilde W_{\mu\nu}$ in Eq.~\eqref{eq:Wsh} is the fundamental object entering \eqref{hmunubasic}. From it one can derive the asymptotic limit of the metric at the future null infinity and, in $D=4$, we have
\begin{equation}\label{}
	\langle H_{\mu\nu}(x) \rangle \sim \frac{1}{4\pi r}\int_0^\infty \left[
	e^{-i\omega U} \tilde{W}_{\mu\nu}(\omega \,n)  
	+
	e^{i\omega U} \tilde{W}^{\ast}_{\mu\nu}(\omega \,n) 
	\right]
	\frac{d\omega}{2\pi}
\,.
\end{equation}
Here $n^\mu=k^\mu/\omega$ is a null vector such that $-n\cdot t=-1$ where $t^\mu$ is the four-velocity of the detector, while $r$ is the radial distance of the detector from the source and $U$ is the retarded time. The dimensionless metric fluctuation in the asymptotic limit is thus given by
\begin{equation}\label{WaveformPMfreqintegral}
	h_{\mu\nu}(x) \sim \frac{4G}{r} 
	\int_{0}^{\infty} 
	e^{-i\omega U}\, \frac{\tilde{W}_{\mu\nu}(\omega \,n)}{\sqrt{8\pi G}}\, \frac{d\omega}{2\pi} + (\text{c.c.})\,,
\end{equation}
where ``c.c.'' stands for complex conjugate.

The leading contribution to~\eqref{eq:Wsh} is of order $G^{3/2}$: it is obtained by using the second line of~\eqref{eq:outin2} to rewrite the bra-vector in~\eqref{eq:Wsh}, focusing on $T$ in $S=1+iT$ and on $1$ in $S^\dagger$. This is given by a tree-level amplitude, so we have
\begin{equation}
  \label{eq:A0W0}
  \tilde{W}_0^{\mu\nu} = \operatorname{FT}[\mathcal A_0^{\mu\nu} ]\,,
\end{equation}
where $\mathcal{A}_0^{\mu\nu}$ is now the amplitude with an extra graviton in the final state with respect to the elastic case discussed in the previous section. For our purposes it is useful to notice that it enjoys a classical expansion similar to the elastic amplitude~\eqref{A04pt}
\begin{equation}\label{A05pt}
	\mathcal A_{0}(p_1,p_2;q_1,q_2)
	=
	\begin{gathered}
		\begin{tikzpicture}
			\path [draw, ultra thick, blue] (-4,2)--(-2,2);
			\path [draw, ultra thick, green!60!black] (-4,1)--(-2,1);
			\path [draw, red] (-3,1.5)--(-2,1.5);
			\filldraw[black!20!white, thick] (-3,1.5) ellipse (.5 and .8);
			\draw[thick] (-3,1.5) ellipse (.5 and .8);
			\node at (-4,2)[left]{$p_1$};
			\node at (-4,1)[left]{$p_2$};
			\node at (-2,1)[right]{$p_3$};
			\node at (-2,2)[right]{$p_4$};
			\node at (-2,1.5)[right]{$k$};
		\end{tikzpicture}
	\end{gathered}
=
\mathcal A_0^{[-2]}(\bar p_1, \bar p_2;q_1,q_2)+\mathcal O(\lambda^0)\,,
\end{equation}
with no corrections scaling as $\mathcal O(\lambda^{-1})$. The explicit result for $\tilde{W}_0^{\mu\nu}$ was already discussed long ago in~\cite{Kovacs:1978eu} and has more recently been rederived and simplified in \cite{Jakobsen:2021smu,Mougiakakos:2021ckm} (see also~\cite{Jakobsen:2021lvp,Mougiakakos:2022sic,Jakobsen:2022psy,DeAngelis:2023lvf,Brandhuber:2023hhl,Aoude:2023dui} for generalizations which include spin or tidal effects).

It was pointed out in~\cite{Caron-Huot:2023vxl} that at subleading PM order a naive generalization of~\eqref{eq:A0W0}, where the classical part of the 5-point 1-loop amplitude is used on the r.h.s., is not correct, in the sense that this is not the result that follows from the KMOC prescription~\eqref{hmunubasic}. We are still going to present the calculation as done historically, by first focusing on the 5-point 1-loop amplitude $\mathcal{A}_1^{\mu\nu}$, as done in~\cite{Brandhuber:2023hhy,Herderschee:2023fxh,Elkhidir:2023dco,Georgoudis:2023lgf}, and then analyze separately the new contribution needed to reconstruct the NLO PM waveform.

\subsection{The 5-point 1-loop amplitude}
\label{ssec:5pt1loop}

The one-loop amplitude $\mathcal A_1^{\mu\nu}$, describing the scattering of the massive states with a graviton in the final state, has a real part $\operatorname{Re}\mathcal A_1^{\mu\nu} = \mathcal B_1^{\mu\nu}$ as well as an imaginary part dictated by its unitarity cuts. We will indicate these contributions as follows
\begin{equation}\label{structureABsscc}
	\mathcal A_1^{\mu\nu} = \mathcal B_1^{\mu\nu} + \frac{i}{2}
	( s^{\mu\nu} + s'^{\mu\nu}) +\frac{i}{2}(c_1^{\mu\nu} + c_2^{\mu\nu} )
\end{equation}
where the contributions related to the cuts are
\begin{equation}\label{Schannel}
	s=
	\begin{gathered}
		\begin{tikzpicture}
			\path [draw, ultra thick, blue] (-4,2)--(-.3,2);
			\path [draw, ultra thick, green!60!black] (-4,1)--(-.3,1);
			\path [draw, red] (-1,1.5)--(-.32,1.5);
			\filldraw[black!20!white, thick] (-3,1.5) ellipse (.5 and .8);
			\draw[thick] (-3,1.5) ellipse (.5 and .8);
			\filldraw[black!20!white, thick] (-1.3,1.5) ellipse (.5 and .8);
			\draw[thick] (-1.3,1.5) ellipse (.5 and .8);
		\end{tikzpicture}
	\end{gathered}
	\qquad\quad
	s'=
	\begin{gathered}
		\begin{tikzpicture}
			\path [draw, ultra thick, blue] (-4,2)--(-.3,2);
			\path [draw, ultra thick, green!60!black] (-4,1)--(-.3,1);
			\path [draw, red] (-3,1.5)--(-2.1,1.5);
			\filldraw[black!20!white, thick] (-3,1.5) ellipse (.5 and .8);
			\draw[thick] (-3,1.5) ellipse (.5 and .8);
			\filldraw[black!20!white, thick] (-1.3,1.5) ellipse (.5 and .8);
			\draw[thick] (-1.3,1.5) ellipse (.5 and .8);
		\end{tikzpicture}
	\end{gathered}
\end{equation}
and
\begin{equation}\label{Cchannel}
	c_1=
	\begin{gathered}
		\begin{tikzpicture}
			\path [draw, ultra thick, blue] (-4,2)--(-.3,2);
			\path [draw, ultra thick, green!60!black] (-4,1)--(-2.1,1);
			\path [draw, red] (-3,1.5)--(-.3,1.5);
			\filldraw[black!20!white, thick] (-3,1.5) ellipse (.5 and .8);
			\draw[thick] (-3,1.5) ellipse (.5 and .8);
			\filldraw[black!20!white, thick] (-1.3,1.75) ellipse (.45 and .55);
			\draw[thick] (-1.3,1.75) ellipse (.45 and .55);
		\end{tikzpicture}
	\end{gathered}
	\qquad\quad
	c_2=
	\begin{gathered}
		\begin{tikzpicture}
			\path [draw, ultra thick, blue] (-4,2)--(-2.1,2);
			\path [draw, ultra thick, green!60!black] (-4,1)--(-.3,1);
			\path [draw, red] (-3,1.5)--(-.3,1.5);
			\filldraw[black!20!white, thick] (-3,1.5) ellipse (.5 and .8);
			\draw[thick] (-3,1.5) ellipse (.5 and .8);
			\filldraw[black!20!white, thick] (-1.3,1.25) ellipse (.45 and .55);
			\draw[thick] (-1.3,1.25) ellipse (.45 and .55);
		\end{tikzpicture}
	\end{gathered}
\end{equation}
and they all possess an infrared (IR) divergent contribution. Instead the ``irreducible'' part $\mathcal B_1^{\mu\nu}$ is real and finite.\footnote{As discussed in \cite{Brandhuber:2023hhy,Herderschee:2023fxh,Elkhidir:2023dco,Georgoudis:2023lgf}, IR divergences also appear in the real part \cite{Weinberg:1965nx} but only in quantum terms.}

In the regime~\eqref{eq:lali}, the classical contribution scales as ${\mathcal O}(\lambda^{-1})$ and we have
\begin{equation}\label{s+}
	\frac{i}{2}(s^{\mu\nu}+s'^{\mu\nu}) \sim i s_+^{[-2]\mu\nu} + \mathcal O(\lambda^0)\,,
\end{equation}
while 
\begin{align}
	\frac{i}{2}\,c^{\mu\nu}_{1,2} \sim \frac{i}{2} \, c_{1,2}^{[-1]\mu\nu} &\sim \mathcal O(\lambda^{-1})\,,\\
	\mathcal B_1^{\mu\nu} = \mathcal B_1^{[-1]\mu\nu}  
	& \sim \mathcal O(\lambda^{-1})\,.
\end{align}
Due to the absence of classical $\mathcal O (\lambda^{-1})$ terms in \eqref{s+}, the leading order of the amplitude is given by the sum of the $S$-channel cuts and should disappear in classical observables, while its classical part is given by the real part and by the $C$-channel cuts~\eqref{structures+}
\begin{equation}\label{structures+}
	\mathcal A_1^{\mu\nu} \sim i s_+^{[-2]\mu\nu} + \left[
	\mathcal B_1^{[-1]\mu\nu}
	+
	\frac{i}{2} \left(
	c_1^{[-1]\mu\nu} + c_2^{[-1]\mu\nu}
	\right)
	\right]
	+\mathcal O(\lambda^{0})\,.
\end{equation}  

The rest of this section is devoted to a closer analysis of the various building blocks of the amplitude. 
Starting from $\mathcal B_1^{\mu\nu}$, we recall that it can be further broken down as follows
\begin{equation}\label{B1OB1E}
	\mathcal B_1^{\mu\nu} = \mathcal B_{1O}^{\mu\nu} + \mathcal B_{1E}^{\mu\nu}\,,
\end{equation}
where 
\begin{equation}\label{B1oddgeneral}
	\mathcal B_{1O}^{\mu\nu} = \left[
	1- \frac{\sigma(\sigma^2-\frac{3}{2})}{(\sigma^2-1)^{3/2}}
	\right]
	\pi G E \omega \, \mathcal A_0^{\mu\nu}
\end{equation}
is \emph{odd} under $\omega_{1}\mapsto - \omega_{1}$, $\omega_{2}\mapsto - \omega_{2}$, $\omega\mapsto - \omega$, while $\mathcal B_{1E}^{\mu\nu}$ is \emph{even} under this transformation. The even part takes the following form,
\begin{equation}\label{B1Efunct}
	\varepsilon_\mu\mathcal B_{1E}^{\mu\nu}\varepsilon_\nu
	=
	\frac{A_{1}^{R}}{\sqrt{q_1^2+\omega_2^2}}
	+
	\frac{A_{2}^{R}}{\sqrt{q_2^2+\omega_1^2}}
	+
	\frac{A_{3}^{R}}{\sqrt{q_1^2}}
	+
	\frac{A_{4}^{R}}{\sqrt{q_2^2}}
\end{equation}
where $A_i^{R}$ are rational functions of $q_1^2$, $q_2^2$, $\omega_1$, $\omega_2$ and $\sigma$, bilinear in $\varepsilon\cdot q_2$, $\varepsilon\cdot u_1$, $\varepsilon \cdot u_2$, which are provided in the ancillary files of \cite{Herderschee:2023fxh}.

Let us turn to the $C$-channel contribution to the waveform kernel \cite{Brandhuber:2023hhy,Herderschee:2023fxh,Elkhidir:2023dco,Georgoudis:2023lgf}.
For these cuts the infrared divergence is associated to a running logarithm depending on the frequency, hallmark of the tail effect, and can be cast in the form,
\begin{equation}\label{tailAM}
	\frac{i}{2} (c_1^{\mu\nu}+c_2^{\mu\nu})
	=
	i G E \omega \left[
	-\frac{1}{\epsilon}
	+
	\log\frac{\omega_1\omega_2}{\mu_\text{IR}^2}
	\right]
	\mathcal A_0^{\mu\nu}
	+i\mathcal M_1^{\mu\nu}
\end{equation}
where $\mathcal M_1^{\mu\nu}$ is IR finite. We follow in particular the splitting adopted in \cite[Eq.~(7.2)]{Herderschee:2023fxh} so that	
$\mathcal M_1^{\mu\nu}$ is given as follows, 
\begin{equation}\label{eq:Ccutfun}
	\begin{split}
	&\varepsilon_\mu \mathcal{M}_1^{\mu\nu} \varepsilon_\nu
	=
	A_{\text{rat}}^{I}
	\\
	&+
	A_{1}^{I}\,
	\frac{\arcsinh\frac{\omega_2}{\sqrt{q_1^2}}}{\sqrt{q_1^2+\omega_2^2}}
	+
	A_{2}^{I}\,
	\frac{\arcsinh\frac{\omega_1}{\sqrt{q_2^2}}}{\sqrt{q_2^2+\omega_1^2}}
	+
	A_{3}^{I}\,
	\log\frac{q_1^2}{q_2^2}
	+
	A_{4}^{I}\,
	\log\frac{\omega_1}{\omega_2}
	+
	A_{5}^{I}\,
	\frac{\arccosh \sigma}{(\sigma^2-1)^{3/2}}\,.
	\end{split}
\end{equation}
$A_i^{I}$ are again rational functions of $q_1^2$, $q_2^2$, $\omega_1$, $\omega_2$ and $\sigma$, bilinear in $\varepsilon\cdot q_2$, $\varepsilon\cdot u_1$, $\varepsilon \cdot u_2$ and are also provided in the ancillary files of \cite{Herderschee:2023fxh}.

For later convenience let us evaluate the divergent part of $c^{\mu\nu}_1$. This  comes from the region where the momentum of the graviton in the Compton sub-diagram (the right blob) is the smallest energy scale in the problem. In this regime we can approximate the explicit expression for this contribution
\begin{equation}\label{}
	\begin{aligned}
	c_{1\mu\nu} &= \begin{gathered}
		\begin{tikzpicture}[scale=.7]
			\path [draw, ultra thick, blue] (-4,2)--(-.3,2);
			\path [draw, ultra thick, green!60!black] (-4,1)--(-2.1,1);
			\path [draw, red] (-3,1.5)--(-.3,1.5);
			\filldraw[black!20!white, thick] (-3,1.5) ellipse (.5 and .8);
			\draw[thick] (-3,1.5) ellipse (.5 and .8);
			\filldraw[black!20!white, thick] (-1.3,1.75) ellipse (.45 and .55);
			\draw[thick] (-1.3,1.75) ellipse (.45 and .55);
		\end{tikzpicture}
	\end{gathered}
=
\int\frac{d^D\ell}{(2\pi)^D}\,2\pi\delta(2\bar p_1\cdot \ell)\,2\pi\delta((k+\ell)^2)\,\Theta(-k^0-\ell^0)
	\\
	&\times \mathcal A_0^{[-2]\rho\sigma}(\bar p_1,\bar p_2; q_1-\ell,q_2)
	\left(
	\delta_\rho^\alpha \delta_\sigma^\beta-
	\frac{1}{D-2}\,
	\eta_{\rho\sigma}\eta^{\alpha\beta}
	\right)
	 \mathcal A^{(C)[-2]}_{\alpha\beta,\mu\nu}(p_1,-\ell-k;\ell)\,,
	\end{aligned}
\end{equation}
by using the GR contribution\footnote{As seen, for instance, by taking the $q_1+q_2\to 0$ limit of Eq.~(4.36) in~\cite{DiVecchia:2023frv}.}
\begin{equation}
  \label{eq:cdivap}
  A^{(C)[-2]}_{\alpha\beta,\mu\nu}(p_1,-\ell-k;\ell)  
  = 
  4 \kappa^2 \frac{(p_1 \cdot k)^2}{\ell^2} 
  \left(
  \eta^{\alpha\mu}-\frac{p_1^{(\alpha}k_{\phantom{1}}^{\mu)}}{p_1\cdot k}
  \right) 
  \left(
  \eta^{\beta\nu}-\frac{p_1^{(\beta}k_{\phantom{1}}^{\nu)}}{p_1\cdot k}
  \right)
  + {\cal O}(\ell)
\end{equation}
and by placing a very small cutoff $\Lambda$ on $|\ell|$, so that we may effectively set $\ell=0$ in $\mathcal{A}_0^{[-2]\rho\sigma}$. Proceeding in this way, and contracting with physical polarizations $\varepsilon\cdot k=0$, $\varepsilon\cdot\varepsilon=0$, we get 
\begin{equation}\label{fordiv}
	\begin{split}
		c_{1\mu\nu}^\text{div}\varepsilon^\mu \varepsilon^\nu
		&= 4 \kappa^2 
		\mathcal A_0^{[-2]\rho\sigma}(\bar p_1,\bar p_2; q_1,q_2)\varepsilon_\rho \varepsilon_\sigma
		\int \frac{d^D\ell}{(2\pi)^D}
		\,2\pi\delta(2p_1\cdot\ell)
		\,2\pi\delta(2k\cdot \ell)
		\Theta(\Lambda^2-\ell^2)\,\frac{(m_1\omega_1)^2}{\ell^2} \,.
	\end{split}
\end{equation}
Using 
\begin{equation}\label{}
	\int \frac{d^D\ell}{(2\pi)^D}\,\frac{1}{\ell^2}
	\,2\pi\delta(2p_1\cdot\ell)
	\,2\pi\delta(2k\cdot \ell)\Theta(\Lambda^2-\ell^2) = \frac{1}{8\pi m_1\omega_1}\left[
	-\frac{1}{2\epsilon}+\log\Lambda
	\right],
\end{equation}
we obtain
\begin{equation}\label{divc1c2}
	\left(c_{1 \mu\nu} + c_{2 \mu\nu} \right)^\text{div}
	=
	2 G E \omega \, \mathcal A_{0 \mu\nu}^{[-2]}(\bar p_1,\bar p_2; q_1,q_2)\,  \left[
	-\frac{1}{\epsilon}+\log(\Lambda^2)
	\right]
\end{equation}
in agreement with the literature~\cite{Brandhuber:2023hhy,Herderschee:2023fxh,Georgoudis:2023lgf,Caron-Huot:2023vxl} (and with \eqref{tailAM}).

\subsection{The NLO PM waveform}
\label{ssec:NLOPMwaveform}

Evaluating at the subleading order~\eqref{eq:Wsh} shows that in order to derive the NLO waveform  the 1-loop amplitude discussed above it is not sufficient to drop the divergent term~\eqref{s+}. Instead one should subtract the cut $s'^{\mu\nu}$~\cite{Caron-Huot:2023vxl}, which yields
\begin{align}\label{W0W1expl}
  \tilde W_1^{\mu\nu} = \operatorname{FT}\left[\mathcal A_1^{\mu\nu} - is'^{\mu\nu}\right] =
  \operatorname{FT}\left[\mathcal B_1^{[-1] \mu\nu} + i s^{[-1]\mu\nu}_- 
  + \frac{i}{2}(c_1^{[-1]\mu\nu}+c_2^{[-1]\mu\nu})\right],
\end{align}
where we define $s_{-}^{[-1]\mu\nu}$ by
\begin{equation}\label{}
	\label{s-}
	\frac{i}{2}(s^{\mu\nu}-s'^{\mu\nu}) \sim i\, s_-^{[-1]\mu\nu} 
		\sim \mathcal O(\lambda^{-1})\,.
\end{equation}
In this section we focus on this new contribution $s^{[-1]\mu\nu}_-$ that was not taken into account in~\cite{Brandhuber:2023hhy,Herderschee:2023fxh,Georgoudis:2023lgf}.

Two approaches are possible. One is to compute $s^{[-1]\mu\nu}_-$ in momentum space by using the explicit expression for the tree-level 4-point and 5-point amplitudes, and then by performing the loop integration, following the same approach used for the amplitude~\cite{Brandhuber:2023hhy,Herderschee:2023fxh,Georgoudis:2023lgf}. We refer to~\cite{Georgoudis:2023ozp,Bohnenblust:2023qmy} and v4 of~\cite{Brandhuber:2023hhy,Herderschee:2023fxh} for this calculation. Here, instead, we follow an alternative approach inspired by the eikonal exponentiation, which allows us to provide a compact relativistic expression for  $\tilde{s}^{[-1]\mu\nu}_-$ directly in impact-parameter space. The idea is simple: instead of first performing the loop integrals and then the Fourier transform~\eqref{eq:ft} to $b$-space, we rewrite the contribution to the cut as a product of {\em tree-level} building blocks in impact parameter space, very much as it was done starting from~\eqref{eq:Q2im} for the cut in the elastic 1-loop amplitude. This approach has the further advantage of eliminating all analytic (in $q_1-q_2$) contributions that translate to short-range terms in $b$-space and are thus irrelevant in the PM expansion, but that make the momentum space expression $s^{[-1]\mu\nu}_-$ more complicated.

We begin from the explicit expressions for the cuts $s$ and $s'$ in terms of the momentum space tree-level ingredients. With an accuracy up to and including the classical $\mathcal O(\lambda^{-1})$ order, we have
\begin{equation}\label{}
	\begin{split}
	s_{\mu\nu}
	&=
	\begin{gathered}
		\begin{tikzpicture}[scale=.7]
			\path [draw, ultra thick, blue] (-4,2)--(-.3,2);
			\path [draw, ultra thick, green!60!black] (-4,1)--(-.3,1);
			\path [draw, red] (-1,1.5)--(-.32,1.5);
			\filldraw[black!20!white, thick] (-3,1.5) ellipse (.5 and .8);
			\draw[thick] (-3,1.5) ellipse (.5 and .8);
			\filldraw[black!20!white, thick] (-1.3,1.5) ellipse (.5 and .8);
			\draw[thick] (-1.3,1.5) ellipse (.5 and .8);
		\end{tikzpicture}
	\end{gathered}=
\int \frac{d^D\ell}{(2\pi)^D}\,2\pi\delta(2\bar p_1\cdot \ell+\ell\cdot(\ell-q_1))\,2\pi\delta(2\bar p_2\cdot \ell-\ell\cdot(\ell+q_2))\\
&\times 
\mathcal A_0^{(4)[-2]}\left(\bar p_1+\tfrac{1}{2}(\ell-q_1),\bar p_2-\tfrac{1}{2}(\ell+q_2);\ell\right)
\mathcal A_{0\mu\nu}^{[-2]}\left(\bar p_1+\tfrac{\ell}{2},\bar p_2-\tfrac{\ell}{2};q_1-\ell,q_2+\ell\right)
\end{split}
\end{equation}
and
\begin{equation}\label{}
	\begin{split}
		s_{\mu\nu}'
		&=
		\begin{gathered}
			\begin{tikzpicture}[scale=.7]
				\path [draw, ultra thick, blue] (-4,2)--(-.3,2);
				\path [draw, ultra thick, green!60!black] (-4,1)--(-.3,1);
				\path [draw, red] (-3,1.5)--(-2.1,1.5);
				\filldraw[black!20!white, thick] (-3,1.5) ellipse (.5 and .8);
				\draw[thick] (-3,1.5) ellipse (.5 and .8);
				\filldraw[black!20!white, thick] (-1.3,1.5) ellipse (.5 and .8);
				\draw[thick] (-1.3,1.5) ellipse (.5 and .8);
			\end{tikzpicture}
		\end{gathered}=
		\int \frac{d^D\ell}{(2\pi)^D}\,2\pi\delta(2\bar p_1\cdot \ell-\ell\cdot(\ell-q_1))\,2\pi\delta(2\bar p_2\cdot \ell+\ell\cdot(\ell+q_2))\\
		&\times
                \mathcal A_{0\mu\nu}^{[-2]}\left(\bar p_1-\tfrac{\ell}{2},\bar p_2+\tfrac{\ell}{2};q_1-\ell,q_2+\ell\right)
		\mathcal A_0^{(4)[-2]}\left(\bar p_1-\tfrac{1}{2}(\ell-q_1),\bar p_2+\tfrac{1}{2}(\ell+q_2);\ell\right).
	\end{split}
\end{equation}
In the sum, we see that all first-order expansions of $\mathcal A_0^{(4)[-2]}$, $\mathcal A_0^{[-2]}$ and of the delta functions in the soft region $\ell\sim q_1\sim q_2\sim \mathcal O(\lambda)$ cancel out and one is left just with the leading $\mathcal O(\lambda^{-2})$ contribution
\begin{equation}\label{}
	\frac{s_{\mu\nu}+s_{\mu\nu}'}{2} 
	=
	\int \frac{d^D\ell}{(2\pi)^{D-2}}\,\delta(2\bar p_1\cdot \ell)\,\delta(2\bar p_2\cdot \ell) 
	\mathcal A_0^{(4)[-2]}\left(\bar p_1,\bar p_2;\ell\right)
	\mathcal A_{0 \mu\nu}^{[-2]}\left(\bar p_1,\bar p_2;q_1-\ell,q_2+\ell\right)
\end{equation}
up to $\mathcal O(\lambda^0)$ corrections, in agreement with \cite{Georgoudis:2023lgf,Caron-Huot:2023vxl}. Vice versa, in the difference they add up. As done in the elastic case we use~\eqref{eq:delpr} and integrate by parts the result of the expansion in order to rewrite the contribution without using derivatives of delta function. Since we are focusing on a quantity whose leading term is classical, we can just neglect all subleading quantum corrections, obtaining
\begin{align}\label{intermediateoperators}
		&\frac{i}{2}\left(
		s^{[-1]}_{\mu\nu}-s^{' [-1]}_{\mu\nu}
		\right)
		=
		i
		\int\frac{d^D\ell}{(2\pi)^D}\,2\pi\delta(2\bar p_1\cdot \ell)\,2\pi\delta(2\bar p_2\cdot \ell)
		\\
		\nonumber
		&
		\times
		\Bigg\{
		\mathcal A_{0 \mu\nu}^{[-2]}(\bar p_1,\bar p_2;q_1-\ell,q_2+\ell)\,
		\frac{E \omega}{2 m_1 m_2}
		\left[
		\frac{\partial}{\partial \sigma}
		-
		\frac{\sigma}{\sigma^2-1}
		\right]
		\mathcal A_0^{(4)[-2]}(\sigma;\ell^2)+
		\mathcal A^{(4)[-2]}(\sigma;\ell^2)
		\\
		\nonumber
		&
		\times
		\left[
		\frac{\ell^2-\ell\cdot q_1}{2 m_1}\frac{\partial}{\partial \ell_{\parallel 1}}
		-
		\frac{\ell^2+\ell\cdot q_2}{2 m_2}\frac{\partial}{\partial \ell_{\parallel 2}}
		+\frac{\ell^\alpha}{2}
		\left(
		\frac{\partial}{\partial \bar p_1^\alpha}
		-
		\frac{\partial}{\partial \bar p_2^\alpha}
		\right)
		\right]
		\mathcal A_{0 \mu\nu}^{[-2]}(\bar p_1,\bar p_2;q_1-\ell,q_2+\ell)
		\Bigg\},
\end{align}
where we have defined $\ell^\mu = \ell_{\parallel 1} \check u_1^\mu + \ell_{\parallel 2} \check u_2^\mu +\ell_\perp^\mu$ and
\begin{equation}\label{}
	\mathcal A^{(4)[-2]}(\bar p_1,\bar p_2;\ell) \simeq \mathcal A^{(4)[-2]}(p_1,p_2;\ell) = \mathcal A^{(4)[-2]}(\sigma;\ell^2)\,.
\end{equation}
Note that the differential operator in the square parenthesis in the last line of \eqref{intermediateoperators} preserves the mass-shell constraints $\bar p_1\cdot (q_1-\ell)=0$ and $\bar p_2\cdot(q_2+\ell)=0$.
One can evaluate the first term of the curly parenthesis in~\eqref{intermediateoperators} by using the elastic $2\to2$ amplitude \eqref{4ptGR}, obtaining
\begin{equation}\label{}
	\frac{1}{2 m_1 m_2}
	\left[
	\frac{\partial}{\partial \sigma}
	-
	\frac{\sigma}{\sigma^2-1}
	\right]
	\mathcal A_0^{(4)[-2]}(\sigma;\ell^2)=
	\frac{16\pi G m_1 m_2}{\ell^2}\,
	\frac{\sigma\left(\sigma^2-\frac{3-4\epsilon}{2-2\epsilon}\right)}{\sigma^2-1}\,.
\end{equation}
We now take the Fourier transform as we did in \eqref{eq:ft} for the elastic amplitude. In the case at hand we solve $q_1+q_2+k=0$ in \eqref{eq:kc5} by eliminating $q_2$ and define the Fourier transform with respect to $q_1$
\begin{equation}\label{FT5b}
	\operatorname{FT}_{q_1}^{(k)}[f(q_1,q_2,k)]
	=
	e^{-ib_2\cdot k}
	\int \frac{d^Dq_1}{(2\pi)^D}\,2\pi\delta(2\bar p_1\cdot q_1)\,2\pi\delta(2\bar p_2\cdot(q_1+k))\,e^{ib\cdot q_1}f(q_1,-q_1-k,k)\,.
\end{equation}
We proceed by considering separately the two terms in \eqref{intermediateoperators}, writing
\begin{equation}\label{}
  i\,\tilde{s}^{[-1]\mu\nu}_-
	=
	\tilde{\mathcal{D}}^{\mu\nu}
	+
	\tilde{\mathcal{F}}^{\mu\nu}\,.
\end{equation}
The $\mathcal{D}$-term factorizes straightforwardly as follows
\begin{equation}\label{Dterm}
	\tilde{\mathcal{D}}^{\mu\nu}
	=
	\frac{i}{2}\,\omega\,
	T_{\rm 1PM}\,\tilde W^{\mu\nu}_0(p_1,p_2;b_J;k)
\end{equation}
in terms of the tree-level waveform $\tilde W_0$ and of the IR-divergent Shapiro time delay~\eqref{TGR}. As in the elastic case, the classical results obtained  by following the KMOC prescription are written in the frame aligned with $p_i$ and $b_J$.
The $\mathcal F$-term can be instead cast in the form
\begin{equation}\label{tildeF}
	\begin{aligned}
		\tilde{\mathcal{F}}_{\mu\nu}
		&=
		Q_{\rm 1PM}^\alpha \left[
		\frac{\check u_1^\beta}{2 m_1}\left(
		-i\frac{\partial}{\partial b_J^\alpha}
		\right)
		+
		\frac{\check u_2^\beta}{2 m_2}\left(
		-i\frac{\partial}{\partial b_J^\alpha} + k_\alpha
		\right)
		\right]
		\operatorname{FT}_\ell^{(k)}
		\left[
		\frac{\partial}{\partial \ell^\beta}\mathcal A_{0\mu\nu}^{[-2]}(\bar p_1, \bar p_2; \ell; k)
		\right]
		\\
		&+
		Q_{\rm 1PM}^\alpha \operatorname{FT}_\ell^{(k)}
		\left[
		\frac12\left(
		\frac{\partial}{\partial \bar p_1^\alpha}
		-
		\frac{\partial}{\partial \bar p_2^\alpha}
		\right)\mathcal A_{0 \mu\nu}^{[-2]}(\bar p_1, \bar p_2; \ell; k)
		\right],
	\end{aligned}
\end{equation}
where we used the 1PM impulse~\eqref{TGR}. In the following, we restrict to the center-of-mass translations frames,
\begin{equation}\label{bicom}
	b_1^{\alpha} = \frac{E_2}{E}  b_J^\alpha\,,
	\qquad
	b_2^\alpha = -\frac{E_1}{E} \,b_J^\alpha\,,
\end{equation}
where 
\begin{equation}\label{CMalphaa}
\frac{E_1}{E} = \frac{m_1(m_1+m_2\sigma)}{m_1^2+2m_1m_2\sigma+m_2^2}\,,
	\qquad
\frac{E_2}{E} = \frac{m_2(m_2+m_1\sigma)}{m_1^2+2m_1m_2\sigma+m_2^2}\,.
      \end{equation}
Following the example of the impulse kernel discussed in the previous section, let us compare the $\mathcal F$-term in \eqref{tildeF} with the transformation of the leading waveform, 
\begin{equation}\label{transformationDEF}
	\bar\delta \tilde W_0^{\mu\nu}
	=
	\tilde W_0^{\mu\nu}(\tilde p_1,\tilde p_2;b_e; k)-\tilde W_0^{\mu\nu}(- p_1,- p_2;b_J; k)\,,
\end{equation}
with $\bar p_1$, $\bar p_2$ and $b_e$ as in \eqref{barp1barp2} and \eqref{betransf}.
To leading PM order, the transformation \eqref{transformationDEF} is given by Taylor expanding the waveform, which leads us to
\begin{equation}\label{}
	\bar \delta \tilde{W}^{\mu\nu} = Q_{\rm 1PM} \,\bar\partial \tilde{W}^{\mu\nu}_0(m_1 v_1, m_2 v_2; b_J; k) + \mathcal O(G^{7/2})
\end{equation}
with $\bar\partial$ as in \eqref{myDiffOp}.
Letting $\bar \partial$ act on the leading-order waveform expressed as the Fourier transform \eqref{FT5b} of the tree-level amplitude $\mathcal A_0^{[-2]}$ we obtain 
\begin{equation}\label{frombardeltatomathcalF}
	\bar\delta \tilde{W}^{\mu\nu}_0 = \tilde{\mathcal{F}}- \frac{i}{2}\, \omega
	b_e \,Q_{\rm 1PM}\,
	\frac{E_1E_2}{E p^2}\,
	\tilde W_0^{\mu\nu} \,,
\end{equation}
where we write $b_e$ at the place of $b_J$ when the difference is immaterial for our purposes and
\begin{equation}\label{}
	E = \sqrt{m_1^2+2 m_1 m_2\sigma +m_2^2}\,,\qquad E p = m_1 m_2 \sqrt{\sigma^2-1}\,.
\end{equation}

Combining \eqref{frombardeltatomathcalF}, from which we read off the $\mathcal F$-term, with the expression for the $\mathcal D$-term in \eqref{Dterm},
we finally obtain
\begin{equation}\label{CHCM}
  i\, \tilde{s}^{[-1]\mu\nu}_-
	=
	\bar\delta \tilde W_0^{\mu\nu}
	+\frac{i}{2}\,
	\omega \left(
	T_\text{1PM}
	+
	b_e \,Q_{\rm 1PM}\,
	\frac{E_1E_2}{E p^2}
	\right)\tilde W_0^{\mu\nu}\,.
\end{equation}
By using the GR expressions for $T_\text{1PM}$ and $Q_\text{1PM}$~\eqref{TGR}, we have the explicit expression
\begin{align}\label{CHsolved}
		&i\,\tilde{s}^{[-1]\mu\nu}_- 		=
		\bar\delta \tilde W_0^{\mu\nu}
		\\
		&+ 
		i G \mu_\text{IR}^{2\epsilon} E \omega
		\left[
		\frac{\sigma \left(\sigma^2-\frac{3-4\epsilon}{2-2\epsilon}\right)}{(\sigma^2-1)^{3/2}} \frac{\Gamma(-\epsilon)}{(\pi b_e^2)^{-\epsilon}}
		 +
		\frac{(m_1+m_2\sigma)(m_2+m_1\sigma)}{m_1^2+2m_1m_2\sigma+m_2^2}\frac{2\sigma^2-1}{(\sigma^2-1)^{3/2}}\right] \tilde W_0^{\mu\nu}
		\,.
\nonumber
\end{align}
The first line of Eq.~\eqref{CHsolved} thus embodies the mechanism showcased by Eq.~\eqref{s-change} for the case of the elastic impulse. KMOC subtractions of unitarity cuts with two intermediate on-shell massive particles both subtract classically singular terms from the amplitude and leave behind the transformation associated to the change of frame \eqref{barp1barp2} and \eqref{betransf}, induced by the operator $\bar\partial$ \eqref{myDiffOp}.
In this sense, \eqref{CHsolved} provides the generalization for generic velocity and frequency of the mechanism first noted in Ref.~\cite{Bini:2023fiz}, in the PN limit, and in Ref.~\cite{Aoude:2023dui} to leading order in the soft limit.

A novelty with respect to the elastic case is instead given by the IR divergent piece in the second line of \eqref{CHsolved}. The IR pole comes from the first term of the square parenthesis there and takes the form
\begin{equation}\label{divergences-cs}
	i	\,\tilde{s}^{[-1]\mu\nu}_{-} 
	=
	-	\frac{i}{\epsilon} 	GE \omega\, \frac{\sigma(\sigma^2-\frac{3}{2})}{(\sigma^2-1)^{3/2}}
	\tilde W_0^{\mu\nu} + \mathcal O(\epsilon^0)\,.
\end{equation}
We can read the IR divergent contribution of the other cuts from~\eqref{divc1c2}: since the dependence on the impact parameter is entirely in $\tilde W_0^{\mu\nu}$, it is straightforward to perform the Fourier transform back to momentum space obtaining
\begin{equation}\label{divtilc}
	\tilde c^{\mu\nu}_1+\tilde c^{\mu\nu}_2 
	=
	\left(\tilde c^{\mu\nu}_1+\tilde c^{\mu\nu}_2 \right)_\text{reg} + \left(\tilde c^{\mu\nu}_1+\tilde c^{\mu\nu}_2 \right)_\text{div}\,,
	\qquad
	\frac{i}{2} \left(\tilde c^{\mu\nu}_1+\tilde c^{\mu\nu}_2 \right)_\text{div}
	=
	-\frac{i}{\epsilon} G E \omega \tilde W^{\mu\nu}_0\,.
\end{equation} 
Since both \eqref{divergences-cs} and \eqref{divtilc} are proportional to the tree-level waveform, one can always reabsorb them into the definition of the origin of retarded time in \eqref{WaveformPMfreqintegral} \cite{Goldberger:2009qd,Porto:2012as}. This is equivalent to multiplication by an overall phase in frequency domain,
\begin{equation}\label{WaveformPMfreqintegral0}
	h_{\mu\nu}(x) \sim \frac{4G}{r} 
	\int_{0}^{\infty} 
	e^{-i\omega (U+\frac{1}{\epsilon}\,U_0 )}\, \frac{\tilde{W}^\text{sub}_{\mu\nu}(\omega \,n)}{\kappa}\, \frac{d\omega}{2\pi} + (\text{c.c.})\,,
\end{equation}
where
\begin{equation}
	\label{eq:Wsub}
	U_0 = G E \left[1+ \frac{\sigma(\sigma^2-\frac{3}{2})}{(\sigma^2-1)^{3/2}} \right], 
\end{equation}
so $\tilde{W}^\text{sub}_{\mu\nu}$ is finite up to NLO in the PM expansion.
In fact also the finite piece in the second line of \eqref{CHsolved} bears this form and can be canceled or modified in a similar way.
By following the same logic, it is convenient to subtract with an appropriate choice of the time-origin the whole second line of~\eqref{CHsolved}, while the first line can be effectively taken into account by using everywhere $b_e$, $\tilde u_1$, $\tilde u_2$, instead of $b_J$, $v_1$, $v_2$ (see Eqs.~\eqref{barp1barp2}, \eqref{betransf}).
This will simplify the formulas presented in Sections \ref{sec:Soft} and \ref{sec:compPN} below.

Let us comment on the comparison between the KMOC subtraction \eqref{W0W1expl} and an analogous one discussed in \cite{Ciafaloni:2015xsr,Ciafaloni:2018uwe} in the ultrarelativistic/massless case. There, $i s^{\mu\nu}$ was subtracted, instead of $i s'^{\mu\nu}$. This leads to a relative minus sign in the analog of \eqref{eq:Wsub}, which explains the absence of IR divergences in that setup as $\sigma\to\infty$.\footnote{We are grateful to Gabriele Veneziano for pointing this out to us.}

We have performed the following cross checks on the above derivation.
First, we have explicitly calculated
both  sides of \eqref{frombardeltatomathcalF} to
leading order in the PN expansion $\omega\sim p_\infty=\sqrt{\sigma^2-1}\to0$. This can be conveniently done by starting from the momentum-space expression of the tree-level 5-point amplitude, expanding it to leading order in the PN regime, and then substituting it into both \eqref{tildeF} and \eqref{transformationDEF}. 
As a separate check, 
we started from the momentum-space expression for the cut in Ref.~\cite{Georgoudis:2023ozp}, expanded it to LO, NLO and NNLO in the PN limit, and performed the resulting Fourier transform to impact-parameter space, finding perfect agreement with Eq.~\eqref{CHsolved}.\footnote{In this check one needs to be careful in two steps. First, one has to use the same IR regulator on both sides of \eqref{CHsolved}; for instance, when using the result of Ref.~\cite{Georgoudis:2023ozp} in the l.h.s. of \eqref{CHsolved}, one needs to use $\bar\mu^{2} = \mu_\text{IR}^{2} 4\pi e^{-\gamma}$ (see Eq.~(3.8) of \cite{Georgoudis:2023lgf}). Second, as clear from its derivation, \eqref{CHsolved} has to be evaluated in $D=4-2\epsilon$ dimensions; there are nontrivial $\epsilon/\epsilon$ terms arising in the $\epsilon\to 0$ limit of the product $T_\text{1PM} \tilde W_0^{\mu\nu}$ on the r.h.s. of~\eqref{CHsolved}, which need to be taken into account.}
The Fourier transforms appearing in the PN limit can be easily computed via
\begin{equation}\label{FTwithomegapinf}
	\int \frac{d^{2-2\epsilon}q_\perp}{(2\pi)^{2-2\epsilon}}
	 \left(
	1+\frac{p_\infty^2q_\perp^2}{\omega^2}
	\right)^\nu
	e^{ib\cdot q_\perp}
	=
	\frac{2^{\nu+\epsilon}}{\pi^{1-\epsilon}}\left(
	\frac{p_\infty}{\omega}
	\right)^{-1+\nu+\epsilon} \frac{K_{1+\nu-\epsilon}\left(\frac{\omega b}{p_\infty}\right)}{\Gamma(-\nu)\,b^{1+\nu-\epsilon}}
\end{equation}
and derivatives thereof.

We have also checked  \eqref{frombardeltatomathcalF} in the soft limit $\omega\to0$ (for generic velocities) by explicitly calculating both of its sides at order $\mathcal O(1/\omega)$, $\mathcal O(\log\omega)$, $\mathcal O(\omega^0)$ and $\mathcal O(\omega\log\omega)$, using the techniques detailed in subsection~\ref{ssec:softintegrals}. 
Moreover, we started from the momentum-space expression for the cut in Ref.~\cite{Georgoudis:2023ozp}, performed the soft limit and checked the agreement in impact parameter space directly with \eqref{CHsolved} to order $\mathcal O(1/\omega)$, $\mathcal O(\log\omega)$, and $\mathcal O(\omega\log\omega)$.
For these calculations, as we will discuss, one can again start from the tree-level 5-point amplitude, but care must be exerted when expanding for small $\omega$ as two regions open up in the Fourier integral and both need to be taken into account.

\section{Soft Expansion}
\label{sec:Soft}

The soft expansion for low graviton frequency of the waveform, or equivalently its early/late limiting behavior domain, is fixed at the first few leading orders by soft theorems \cite{Weinberg:1964ew,Weinberg:1965nx,Strominger:2014pwa,Sahoo:2018lxl,Saha:2019tub,Sahoo:2021ctw}. These universal terms can be obtained by acting with an appropriate soft operator on the elastic eikonal $e^{2i\delta}$~\cite{FrancescoPaolo}. It is therefore interesting to exhibit explicitly the soft expansion of the waveform up to one loop.

The expansion of the waveform as $\omega\to0$ takes the following form,
\begin{equation}\label{softexpasion}
	\tilde{W} = \tilde{W}^{[\omega^{-1}]} + \tilde{W}^{[\log\omega]} + \tilde{W}^{[\omega^0]} + \tilde{W}^{[\omega(\log\omega)^{2}]}
	+
	\tilde{W}^{[\omega\log\omega]}+\cdots
\end{equation}
and the non-analytic terms are constrained by classical soft theorems \cite{Laddha:2018vbn,Sahoo:2018lxl,Sahoo:2021ctw}. The last term displayed in \eqref{softexpasion} is dictated by the soft theorem \cite{Ghosh:2021bam} only at tree level. 

\subsection{Method of regions in the soft limit}
\label{ssec:softintegrals}
Although the $\omega\to0$ limit can be in principle performed on the full $b$-space expression of the waveform, it is often more convenient to anticipate it and take $\omega$ to be small before performing the Fourier transform from momentum space to impact parameter. When doing so, one needs to consider two regions.

The first region is defined in terms of the following kinematic limit, 
\begin{equation}\label{region1}
	\omega \ll q_\perp \sim b^{-1} 
\end{equation}
with
\begin{equation}\label{}
	k^\mu = \omega\, n^\mu\,,
	\qquad
	\omega_1 = \omega\, \alpha_1\,,
	\qquad
	\omega_2 = \omega\, \alpha_2\,,
	\qquad
	q_1^\mu = - \omega\,\alpha_2 \,\check u_2^\mu + q_\perp^\mu\,.
\end{equation} 
This region leads to dramatic simplification of the integrand to be Fourier-transformed, especially at one-loop level. There, by a suitable use of the Gram determinant identity that guarantees the cancellation of spurious poles, we find that all Fourier transforms can be reduced to the following elementary one, which can be seen as a limiting case of \eqref{FTwithomegapinf},
\begin{equation}\label{basicFT}
	\int \frac{d^{2-2\epsilon}q_\perp}{(2\pi)^{2-2\epsilon}}\,(q_\perp^2)^{\nu} \,e^{ib\cdot q_\perp}
	= \frac{4^\nu}{\pi^{1-\epsilon}}\frac{\Gamma(1+\nu-\epsilon)}{\Gamma(-\nu)(b^2)^{1+\nu-\epsilon}}\,.
\end{equation}
Contributions arising from this region are non-analytic in $q_\perp^2$ and therefore long-range in $b$, while they are analytic in $\omega$.

The second region is instead characterized by
\begin{equation}\label{region2}
	\omega \sim q_\perp \ll b^{-1}\,.
\end{equation}
In this region, the integrand does not simplify, since by construction it is homogeneous in $\omega \sim q_\perp$ in the classical limit, but one may expand out the phase factor $e^{ib\cdot q_\perp} = 1+ib\cdot q_\perp+\cdots$. The Fourier integral then effectively turns into an ordinary one. For the tree level, calculations in this region can be systematized by introducing the integral family
\begin{equation}\label{}
	I_{i_1i_2} = \int \frac{d^{2-2\epsilon}q_\perp}{(2\pi)^{2-2\epsilon}}\frac{1}{\left(q_\perp^2+\frac{\omega^2\alpha_2^2}{\sigma^2-1}\right)^{i_1}
		\left((q_\perp-n_\perp)^2+\frac{\omega^2\alpha_1^2}{\sigma^2-1}\right)^{i_2}}
\end{equation}
(the subscript $\perp$ stands for projection orthogonal to $u_1$, $u_2$)
with master integrals
\begin{align}
	I_{10} = \frac{\Gamma(\epsilon)}{(4\pi)^{1-\epsilon}}\,
	\left(
	\frac{\alpha_2^2\omega^2}{\sigma^2-1}
	\right)^{-\epsilon},\qquad
	I_{01} = \frac{\Gamma(\epsilon)}{(4\pi)^{1-\epsilon}}\,
	\left(
	\frac{\alpha_1^2\omega^2}{\sigma^2-1}
	\right)^{-\epsilon}
\end{align}
and
\begin{equation}\label{}
	I_{11} = \frac{\sqrt{\sigma^2-1}}{4\pi\alpha_1\alpha_2\omega^2}
        \arccosh \sigma
        +\mathcal O(\epsilon)\,.
\end{equation}
Contributions arising from this region are instead analytic in $q_\perp^2$, but non-analytic in $\omega$.

For later convenience, it is useful to introduce also the quantities
  \begin{equation}
    \label{eq:baralpha}
    \tilde\alpha_1=-\tilde u_1\cdot k\,, \qquad \tilde \alpha_2 =-\tilde u_2\cdot k\,,
  \end{equation}
which are adapted to the eikonal reference frame. When writing the NLO PM waveform in this frame, which as already mentioned takes into account in a simple way the first term of~\eqref{CHsolved}, one has to use the quantities~\eqref{eq:baralpha} in all tree-level terms.

\subsection{Universal soft terms $1/\omega$, $\log\omega$ and $\omega(\log\omega)^2$}
\label{ssec:softuniversal}

To leading order in the soft limit, we find that the amplitude-based waveform matches the prediction of the leading soft graviton theorem, i.e.~the (linear) memory effect,\footnote{As is well known~\cite{Christodoulou:1991cr,Wiseman:1991ss,Thorne:1992sdb}, there are extra contributions to~\eqref{leadingsoft} due to the non-linearities of gravity which we will neglect in this work: in the PM scattering waveform they start at ${\mathcal O}(G^4)$ for the metric fluctuation.}
\begin{equation}\label{leadingsoft}
  {\tilde W^{[\omega^{-1}]}}
  = \frac{i \kappa Q}{b_e\omega \tilde\alpha_1^2\tilde\alpha_2^2}
	(\tilde\alpha_1 \tilde u_2\cdot \varepsilon-\tilde\alpha_2 \tilde u_1\cdot \varepsilon)
	(2\tilde\alpha_1\tilde\alpha_2 b_e\cdot \varepsilon+b_e\cdot n(\tilde\alpha_1 \tilde u_2\cdot \varepsilon+\tilde\alpha_2 \tilde u_1\cdot \varepsilon))\,,
\end{equation}
where $Q = Q_\text{1PM} + Q_\text{2PM} +\cdots $ is the PM impulse~\eqref{QuptoorderG2}. For our purposes is sufficient to keep the NLO PM corrections and so we can stop at $Q_\text{2PM}$ for the impulse and neglect the difference between $\tilde{m}_i$ and $m_i$.
The $\mathcal O(G)$ contribution to \eqref{leadingsoft} is obtained by straightforward expansion of the tree-level amplitude. Its $\mathcal O(G^2)$ arises instead from the leading soft limit of the real part of the one-loop waveform kernel, $\mathcal B_1$, in particular from its even part $\mathcal B_{1E}$ (since the odd part $\mathcal B_{1O}$ is further suppressed in the soft limit, see Eq.~\eqref{B1oddgeneral}).

We also find that the $\log\omega$ soft term of the waveform matches the universal prediction \cite{Sahoo:2018lxl,Sahoo:2021ctw} \begin{equation}\label{logomega}
	\begin{split}
	{\tilde W^{[\log\omega]}}
	&=
	\kappa \frac{2G m_1 m_2\sigma(2\sigma^2-3)}{\tilde\alpha_1\tilde\alpha_2(\sigma^2-1)^{3/2}}\,(\tilde\alpha_1 \tilde u_2\cdot\varepsilon-\tilde\alpha_2 \tilde u_1\cdot\varepsilon)^2
		\log\left(\frac{\omega b_e\, e^{\gamma}}{2\sqrt{\sigma^2-1}}\right)\\
	&+
	2 i G  E \omega\, {\tilde W_0^{[\omega^{-1}]}}\,\log\omega + {\mathcal O}(G^3)\,.
      \end{split}
\end{equation}
The first line of \eqref{logomega} is the tree-level contribution, for which one does not expect any running logarithm. In order to make this manifest, we made the argument of the logarithm dimensionless by including regular terms that naturally arise from the soft limit of the full tree-level frequency-domain waveform.
Instead the second line of \eqref{logomega} is the one-loop contribution, but it arises entirely from the tail effect displayed explicitly in Eq.~\eqref{tailAM}, and thus it is ultimately dictated by the (soft) tree-level amplitude. The logarithm in the second line of \eqref{logomega} is therefore running, and its scale is set by the dimensional regularization $\mu_\text{IR}$ in our approach. We find that the remainder $\mathcal M_1$ in \eqref{tailAM} does not give $\log\omega$ contributions. Indeed, considering its expansion the region defined by \eqref{region1}, we find that the Fourier transform of its $\mathcal O(\omega^0)$ contributions is finite and does not give rise to any $\log(b)$. Therefore, by dimensional analysis, no $\log(\omega)$ can arise from the other region defined by \eqref{region2}. 

For the $\omega (\log \omega)^2$ behavior, again looking at the terms explicitly displayed in  \eqref{tailAM}, we find
\begin{equation}\label{omegaloglog}
	\tilde W^{[\omega(\log\omega)^2]} = 2i G E \omega {\tilde W_0^{[{\log\omega}]}} \log\omega + {\mathcal O}(G^3)\,,
\end{equation}
once again in perfect agreement with the corresponding soft theorem \cite{Sahoo:2021ctw}. Note that, following the logic highlighted by the comments in the previous paragraph, $\tilde W^{[\omega(\log\omega)^2]}$ in~\eqref{omegaloglog} involves in a natural way the product of a $\log u_{\text{KT}}$, with $u_{\rm KT}=\omega b_e/\sqrt{\sigma^2-1}$ as in~\eqref{eq:pinf} below, and of a running $\log\omega$. 

\subsection{Non-universal soft term $\omega\log\omega$}
\label{ssec:softnonuniversal}

While the tree-level contribution to the $\omega\log\omega$ term,\footnote{Since we are now working in the MPM frame, the eikonal impact parameter must be used throughout and $b_i$ in~\eqref{olotree} should be interpreted as in~\eqref{bicom}, but with $b_e$ at the place of $b_J$.}
\begin{equation}\label{olotree}
	\begin{split}
{\tilde{W}_0^{[\omega\log\omega]}}
&= \kappa
\frac{2 i  G m_1 m_2 \sigma(2\sigma^2-3)}{\tilde\alpha_1\tilde\alpha_2(\sigma^2-1)^{3/2}} 
(\tilde\alpha_1\,\tilde u_2\cdot \varepsilon-\tilde\alpha_2\,\tilde u_1\cdot \varepsilon)
\\
&\times[\tilde\alpha_1\tilde\alpha_2\, b_e\cdot \varepsilon+\tilde\alpha_2 (b_{1}\cdot n)(\tilde u_1\cdot \varepsilon)-\tilde\alpha_1 (b_{2}\cdot n)(\tilde u_2\cdot \varepsilon)]\,\omega\log\omega
	\end{split}
\end{equation}
is completely fixed by the corresponding soft theorem \cite{Ghosh:2021bam}, the one-loop contribution is not, and is sensitive to non-universal terms.

A first one-loop contribution comes from $\mathcal B_{1O}$ in \eqref{B1oddgeneral} and is fixed in terms of a universal tree-level term, 
\begin{equation}\label{eq:B1pi}
	{\tilde{\mathcal{B}}_{1O}^{[\omega\log\omega]}}
	=
	\left[
	1- \frac{\sigma(\sigma^2-\frac{3}{2})}{(\sigma^2-1)^{3/2}}
	\right]
	\pi G E\omega\, \tilde W^{[\log\omega]}_0\,.
\end{equation}
No contributions arise instead from $\mathcal B_{1E}$. We have reached this conclusion by explicitly calculating its expansion up to $\omega^1$ in the region \eqref{region1} and by checking that the Fourier transform \eqref{basicFT} of the resulting expression is finite and does not give rise to any $\log(b)$ (hence, no $\log\omega$ can come from the other region \eqref{region2}). From the $C$-channel cuts, \eqref{tailAM}, we have 
\begin{equation}\label{omegalogomega}
\begin{split}
	\frac{i}{2}(\tilde c_1+\tilde c_2)^{[\omega\log\omega]}
&=
i G E
\left[
-\frac{1}{\epsilon}
+\log\frac{\alpha_1\alpha_2}{\mu_\text{IR}^2}
\right]
\omega \tilde W_0^{[\log\omega]}
\\
&+
2i G E \omega \log\omega\,
 \tilde W_0^{[\omega^0]}
+i \tilde{\mathcal{M}}_1^{[\omega\log\omega]}\,,
\end{split}
\end{equation}
where $\tilde W_0^{[\log\omega]}$ is given in the first line\footnote{At subleading order in the PM expansion, analytic terms in $\omega$ are irrelevant for our analysis of the soft limit, so the equations of these section should be understood up to such terms.} of \eqref{logomega}, while $\tilde W_0^{[{\omega^0}]}$ and $ \tilde{\mathcal{M}}_1^{[\omega\log\omega]}$ will be given in \eqref{omega0} and \eqref{calMomegalogomega} below.
Eq.~\eqref{omegalogomega} involves a universal part in the first line (including the scale of the running logarithm in \eqref{omegaloglog}) and  non-universal contributions in the second line.

The tree-level term $\tilde W_0^{[\omega^0]}$ can be obtained either by expanding the known tree-level waveform \cite{Kovacs:1977uw,Kovacs:1978eu,Jakobsen:2021smu,Mougiakakos:2021ckm} or can be more conveniently calculated by expanding the tree-level amplitude \cite{Luna:2017dtq} in the two regions \eqref{region1}, \eqref{region2} and summing the two contributions. In this second way one sees that, separately, each region gives rise to singularities which however cancel in the sum. Correspondingly, $\log(b)$ and $\log\omega$ terms arising from the two regions neatly combine to reconstruct $\log(\omega b)$ in the sum, thus unambiguously fixing  the non-logarithmic terms as well. 
As a result, we have 
\begin{align}
	\nonumber
	\tilde W_0^{[\omega^0]} & =
                    \kappa (\tilde\alpha_1 \tilde u_2\cdot\varepsilon-\tilde\alpha_2 \tilde u_1\cdot\varepsilon)^2
	\left[\frac{G m_1 m_2\sigma(2\sigma^2-3)}{\tilde\alpha_1\tilde\alpha_2(\sigma^2-1)^{3/2}}\log\left( \tilde\alpha_1\tilde\alpha_2\right)
	-\frac{2G m_1 m_2(2\sigma^2-1)}{\mathcal P \sqrt{\sigma^2-1}}
	\right]
	\\
	\nonumber
	&+
	\frac{4 G m_1 m_2}{\mathcal{P}}
	\Big[
	\frac{(\tilde\alpha_1 \tilde u_2\cdot\varepsilon-\tilde \alpha_2 \tilde u_1\cdot\varepsilon)^2}{\tilde \alpha_1\tilde \alpha_2\mathcal{P}}
	\left(g_3 \arccosh \sigma 
	+g_2\log\frac{\tilde \alpha_1}{\tilde \alpha_2}
	\right)\\
	&
	+
	\frac{2\sigma^2-1}{2b^2\tilde \alpha_1^2\sqrt{\sigma^2-1}}\,g_1
	\Big]
	+
	i b_2\cdot n\,{\omega \tilde W_0^{[\omega^{-1}]}}\,.
	\label{omega0}
\end{align}
Here,
\begin{equation}\label{}
	\mathcal P = -\tilde \alpha_1^2+2\tilde \alpha_1 \tilde \alpha_2 \sigma-\tilde \alpha_2^2\ge0\,,
\end{equation}
and
\begin{subequations}
	\begin{align}
		g_3&=\tilde \alpha _2^2 \tilde \alpha _1^2 \left(2 \sigma ^2+1\right)-2 \tilde \alpha _2 \tilde \alpha _1^3 \sigma -2 \tilde \alpha _2^3 \tilde \alpha _1 \sigma +\tilde \alpha _1^4+\tilde \alpha _2^4\,,\\
		g_2&= -\frac{\tilde \alpha_1^2-\tilde \alpha_2^2}{4(\sigma^2-1)^{3/2}}\left[
		\sigma(2\sigma^2-3) \left(\tilde \alpha _1^2+\tilde \alpha _2^2\right) +2 \tilde \alpha _1 \tilde \alpha _2
		\right],\\
		g_1&=-\mathcal{P} (\tilde \alpha_1\, b_e\cdot\varepsilon+(b_e\cdot n)(\tilde u_1\cdot \varepsilon) )^2 \,.
	\end{align}
\end{subequations}
Note that combining the two terms in the last line of \eqref{omega0} makes the expression manifestly symmetric under particle-interchange symmetry, but we prefer to keep them separate in order to highlight their different origin.

Finally, from the remainder $\mathcal M_1$ in Eq.~\eqref{tailAM}, we obtain 
\begin{equation}\label{calMomegalogomega}
	\begin{split}
	{i\tilde{\mathcal{M}}^{[\omega\log\omega]}_1}
	&= i\kappa
	\omega \log\omega\, G^2\, m_1^2 m_2 \frac{2\sigma (\alpha_1 \, u_2\cdot \varepsilon-\alpha_2 \, u_1\cdot \varepsilon)^2}{(\sigma^2-1)^{3/2}\mathcal{P}}\\
	&\times
	\left[
	\frac{2\sigma^2-3}{\mathcal P}\left(
	f_3\,\frac{\arccosh \sigma}{(\sigma^2-1)^{3/2}}
	+
	f_2\,\frac{1}{\alpha_2}\,\log\frac{\alpha_1}{\alpha_2}
	\right)
	-
	\frac{f_1}{\alpha_2(\sigma^2-1)}
	\right]
	+
	(1\leftrightarrow2)\,,
	\end{split}
\end{equation} 
where we can neglect the difference between tilded and untilded quantities, since this result is already a NLO PM quantity, and
\begin{subequations}
	\begin{align}
		\begin{split}
	f_3&=\alpha _1^3 \left(4 \sigma ^4-6 \sigma ^2+1\right)+\alpha _2 \alpha _1^2 \sigma  \left(-4 \sigma ^4+4 \sigma ^2+3\right)\\
	&+\alpha _2^2 \alpha _1
	\left(4 \sigma ^4-6 \sigma ^2-1\right)+\alpha _2^3 \sigma  \left(3-2 \sigma ^2\right),
	\end{split}
	\\
	f_2&=-\alpha _1^4-\alpha _2^4\,,\\
	f_1&=\alpha _1^2 \left(-\left(\sigma ^2-1\right)\right)+\alpha _2 \alpha _1 \sigma  \left(4 \sigma ^4-6 \sigma ^2+1\right)+\alpha _2^2 \left(-4
	\sigma ^4+7 \sigma ^2-2\right).
	\end{align}
\end{subequations}
Since we are targeting the $\omega \log\omega$ terms, we calculated \eqref{calMomegalogomega} by expanding $\mathcal M_1$ only in the first region, \eqref{region1}, we disregarded the $1/\epsilon$ terms, which would cancel with analogous ones arising from the other region, and focused on the $\omega\log(b)$ terms, which, by dimensional analysis, must appear with the same prefactor as the $\omega\log\omega$ terms in the other region \eqref{region2}. 

Collecting all the relevant contributions, from~\eqref{W0W1expl} and~\eqref{softexpasion}, we can then extract the $\omega \log \omega$ term of the NLO subtracted amplitude~\eqref{WaveformPMfreqintegral0}.  For this term, which will be the main focus of the next section, we have
\begin{equation}
  \label{eq:W1olo}
  \tilde W_1^{[\omega\log\omega]} = \tilde{\mathcal{B}}_{1O}^{[\omega\log\omega]} + \frac{i}{2}(\tilde c_1+\tilde c_2)_{\rm reg}^{[\omega\log\omega]}\,,
\end{equation}
where we used~\eqref{divtilc}, and let us recall that we can neglect all analytic terms in $\omega$ and NNLO PM contributions on the right hand side of \eqref{eq:W1olo}.

\section{Gravitational-wave tails in the soft regime}
\label{sec:compPN}

In the Post-Newtonian (PN) approach it is convenient to expand the metric fluctuation~\eqref{WaveformPMfreqintegral0} at null-infinity in multipole moments of the $SO(3)$ acting on $\hat{n}$, the spatial part of $k^\mu/\omega$, 
\begin{equation}\label{}
	k^\mu= \omega (1,\hat n)\,,
\end{equation}
working in the center-of-mass frame
(see~\cite{Blanchet:2013haa} and references therein). The reason for this is that higher order multipoles are suppressed in small velocity limit,
\begin{equation}
  \label{eq:pinf}
  \sigma=\sqrt{1+p^2_\infty}\,,\qquad 
  p_\infty \ll 1\qquad 
  \text{with }u_\text{KT} = \frac{\omega b_e}{p_\infty}~\text{ fixed} \,.
\end{equation}
To be precise, increasing by one the order of the multipole considered brings an extra factor of $p_\infty$. There are two types of multipoles involved in the decomposition mentioned above which are usually denoted by $U_L$ and $V_L$ which indicate symmetric trace-free tensors of order $\ell$ in Cartesian 3D space. We can reconstruct the spatial part of the PN waveform just by using these multipoles as follows~\cite{Blanchet:2013haa}
\begin{equation}
  \label{eq:WUij}
  {\tilde{W}_{i j}(\omega \,n)} = \sum_{\ell=2}^\infty\frac{\kappa}{\ell!} \left[n^{i_1}\cdots n^{i_{\ell-2}} \mathrm{U}_{ij i_1 \cdots i_{\ell-2}} - \frac{\ell}{\ell+1} n^k n^{i_1}\cdots n^{i_{\ell-2}} \left(\epsilon_{khi}  \mathrm{V}_{j h i_1\ldots i_{\ell-2}} + i \leftrightarrow j\right)\right],
\end{equation}
where we follow the notation of~\cite{Bini:2023fiz}. The physical waveform is the projection of the result above along the physical ``$\times$'' and ``$+$'' polarizations, and have been extensively used in the PN literature to extract more inclusive observables such as the radiated (angular) momentum, see for instance \cite{Bini:2021jmj,Bini:2021qvf,Cho:2021onr,Cho:2022pqy,Bini:2022enm}. Again by following~\cite{Bini:2023fiz}, we will parametrize the result  in terms of the angles of $\hat{n}$ adopting the conventions,
\begin{equation}
  \label{eq:nmbarm}
  \begin{aligned}
    \hat{n}& =(\sin\theta\cos\phi,\sin\theta\sin\phi,\cos\theta)\,,
    \\ 
    \vec{\varepsilon} & = \frac{1}{\sqrt{2}} (\cos\theta \cos\phi + i \sin\phi,\cos\theta \sin\phi - i \cos\phi,-\sin\theta)\,, \\
    \vec{\tilde{p}} & = |\vec p\,|(0,1,0)\;,\quad \vec{b}_e = b_e (1,0,0)\;,
    \end{aligned}
\end{equation}
so that $\hat{n}\cdot \vec{\varepsilon}=\vec{\varepsilon}\cdot \vec{\varepsilon}=0$, by using
\begin{equation}
  \label{eq:WUang}
  \tilde{W}(\omega,\theta,\phi) = {\varepsilon}^i {\varepsilon}^j \tilde{W}_{ij} (\omega \,n)\;.
\end{equation}

The standard approach is to calculate the gravitational field in the near-zone by solving the Einstein equations with a stress-energy tensor appropriate to the binary motion and then analyze its propagation up to future null infinity (again see~\cite{Blanchet:2013haa} for a review). For the tree-level waveform the second part of the calculation is trivial and one can simply identify the $U_L$'s and $V_L$'s with the near-zone field (whose multipoles are usually referred to as $I_L$'s for the representation corresponding to $U_L$ and $J_L$ for the one corresponding to $V_L$). One can obtain these $I_L$'s and $J_L$'s at tree level by simply taking the limit~\eqref{eq:pinf} on the leading PM waveform~\cite{Kovacs:1977uw,Kovacs:1978eu,Jakobsen:2021smu,Mougiakakos:2021ckm} and performing the multipolar decomposition. We checked up to $\ell=4$ that this yields the correct PN results for $U_L$ and $V_L$ as already emphasized in~\cite{Bini:2023fiz}. At subleading PN orders there are several effects that need to be taken into account to reconstruct the full waveform at null infinity starting from the near-zone multipoles. They are neatly summarized in~\cite{Bini:2023fiz}, where they are explicitly spelled out for the scattering case of interest to us. It is interesting to connect the various contributions on the PN side to the different terms of the PM waveform as derived from scattering amplitudes. The following pattern emerges:
\begin{itemize}
\item The expansion in the PN regime~\eqref{eq:pinf} of the tree-level PM waveform yields corrections weighted by $p_\infty^{2k}$, with $k=1,2,\ldots$, to the leading contributions of {\rm each} multipole (so, integer PN corrections of order $k$ relative to the leading term, according to the standard nomenclature). This can be checked explicitly, but also proved in general by noticing that the PM-waveform is even under the transformation
  \begin{equation}
    \label{eq:z2t}
    (p_\infty,\hat{n}) \leftrightarrow (-p_\infty,-\hat{n})
    \qquad 
    \mbox{with $u_\text{KT},b$ fixed}\,.
  \end{equation}
  This means that each factor of $p_\infty$ arising from the expansion is accompanied by a factor of $\hat n$, which changes the order $\ell$ of the multipole by one. For instance, can use the explicit expression of the frequencies $\omega_i$ in the center of mass frame,
      \begin{equation}
        \label{eq:omegacom}
       \begin{aligned}
      \omega_1 = \omega\, \frac{m_1 + m_2 (\sigma - \vec{v} \cdot \hat{n}  \,p_\infty) }{E}
\qquad 
\omega_2 = \omega\, \frac{m_2 + m_1 (\sigma +  \vec{v} \cdot \hat{n} \, p_\infty)}{E}\;
\end{aligned}
\end{equation}
to check that they have the same parity as $\omega$ under this transformation, {\em i.e.}~they are odd. 
\item Also the even part of the irreducible one-loop contribution (dubbed $\mathcal B_{1E}^{\mu\nu}$ in~\eqref{B1OB1E} is even under~\eqref{eq:z2t}. So it has the same parity as the second term in the square parenthesis in~\eqref{B1oddgeneral} and both yield relative PN corrections that have integer order. By taking the PN expansion of these terms, we reproduced Eq.~(7.1) and~(7.2) of~\cite{Bini:2023fiz}, noticing that the first result comes from the part of $\mathcal B_{1O}^{\mu\nu}$, which is even under~\eqref{eq:z2t}, while the second one comes from $\mathcal B_{1E}^{\mu\nu}$.
\item The contribution coming from the $s$-channel cut discussed in Sect.~\ref{ssec:NLOPMwaveform} is again even under~\eqref{eq:z2t}. This is consistent with~\eqref{CHsolved} and thus provides further support for that result.
  \item Instead the contributions coming from the $c$-channel cuts discussed in Sect.~\ref{ssec:5pt1loop} are odd under~\eqref{eq:z2t} exactly as the first term in the square parenthesis in~\eqref{B1oddgeneral}. Thus the PN expansion of these terms does not mix with that of the rest of the NLO PM waveform and yields half-integer PN corrections to the leading contribution for each multipole ({\em i.e.} terms weighted by $p_\infty^{2k+1}$, with $k=1,2,\ldots$).
\end{itemize}

We checked the properties above numerically on the full 1-loop results and it would be interesting to find a more analytic understanding of this structure and, in particular, why the $s$-channel and the $c$-channel cuts behave differently under~\eqref{eq:z2t}. 

In any case, this makes the mismatches at $2.5$ PN order found in~\cite{Bini:2023fiz} even more interesting as they cannot be solved by the inclusion of the new contribution pointed out in~\cite{Caron-Huot:2023vxl} and discussed in Sect.~\ref{ssec:NLOPMwaveform}. As emphasized in~\cite{Bini:2023fiz}, the most surprising mismatch is in the probe limit, where one of the masses is taken to be much larger than the other one, for instance we can take $m_1\gg m_2$. It is standard to introduce
\begin{equation}
  \label{eq:nucr}
  \nu = \frac{m_1 m_2}{m^2}
  \,,\qquad 
  m= m_1+m_2
  \,,\qquad
  \Delta = \frac{m_1-m_2}{m}\,,
\end{equation}
and so in the probe limit the waveform is linear in the symmetric mass-ratio $\nu$. The order $G^2$ of this linear-in-$\nu$ contribution to $\tilde{W}$ is {\em entirely} captured by a very simple result in the PN approach: the so-called tail terms~\cite{Blanchet:1989ki,Blanchet:1992br,Blanchet:1993ec,Blanchet:1995fr,Porto:2012as}. Thus, let us first focus on this contribution ${\cal O}(G^2\nu)$ of the waveform and revisit the mismatch between the PM amplitude-based and PN results which is discussed in~\cite{Bini:2023fiz}. 
We will do so in the soft limit by using the results of Sect.~\ref{sec:Soft}. As already discussed there, the first three terms in the soft expansion of the PM waveform agree with what is obtained from the soft theorems and, as shown in~\cite{Bini:2023fiz}, match the PN results. So from now we will focus on the $\omega\log\omega$ term which is the first one where the mismatch found in~\cite{Bini:2023fiz} appears.

Starting on the PM side, we can consider the two contributions to $\tilde W_1^{[\omega\log\omega]}$ in~\eqref{eq:W1olo} and adjust the  scale $\mu_{\text{IR}}$ by setting 
	\begin{equation}\label{choicemuIR}
		\log\mu_{\text{IR}} = -\log\tilde{b}_0 + \frac{1}{2}\,,
	\end{equation}
for later convenience,\footnote{The $+\frac{1}{2}$ is chosen so as to directly match \eqref{MPMomegalogomegafull} to leading PN order.} obtaining
\begin{equation}
  \label{eq:W1pinopi}
  \tilde W_1^{[\omega\log\omega]} =  \mathcal{B}_{1O}^{[\omega\log\omega]} +  \tilde W_{C}^{[\omega\log\omega]}
  +
  2i G E
  \,\log \tilde b_0\,
  \omega \tilde W_0^{[\log\omega]}
   \;,
\end{equation}
where the first term has an extra factor of $\pi$ with the respect to the second one, which comes instead from the Compton cuts~\eqref{omegalogomega}.  By focusing on the latter, which contains the non-universal part, and saturating it as in~\eqref{eq:WUang}, we obtain,
\begin{align}\label{PNomegalogomegafull}
	\nonumber
	&\frac{\tilde W_{C}^{[\omega\log\omega]}}{\kappa \omega\log\omega}
	=
	-\frac{1}{24\,p_\infty} i G^2 m^3 \nu  \left(35 (\cos (2 \theta )+3) \cos (2 \phi )+140 i \cos (\theta ) \sin (2 \phi )+22 \sin ^2(\theta )\right)
	\\
	\nonumber
	&+
	\frac{1}{60} G^2 m^3 \Delta\,\nu  \sin (\theta ) (\cos (\phi )+i \cos (\theta ) \sin (\phi )) (\cos (\theta ) (307 \cos (2 \phi )-67)+614 i \sin (\phi ) \cos (\phi ))
	\\
	\nonumber
	&+
	\frac{1}{120}\,p_\infty\, i G^2 m^3 \nu  \Big[\Big(139 \cos (2 \theta )+57 \cos (4 \theta )+600 \Big) \cos (2 \phi )\\
	\nonumber
	&+
	2 \sin ^2(\theta ) \Big(\cos (2 \theta ) (51 \cos (4 \phi )+63)+204 i \cos (\theta )
	\sin (4 \phi )+153 \cos (4 \phi )+460\Big)\\
	&+
	4 i \Big(178 \cos (\theta )+21 \cos (3 \theta )\Big) \sin (2 \phi )\Big]
	+\mathcal O(p_\infty^2) + (\text{quadratic in }\nu)\,.
\end{align}

On the PN side, when focusing on the linear-in-$\nu$ sector, we can neglect the non-linearities of gravity and focus just on the near-zone multipoles (the $I_L$'s) and the contribution from the tail formula~\cite{Blanchet:1989ki,Blanchet:1992br,Blanchet:1995fr}\footnote{At the order of interest here, we can neglect the distinction between intermediate- and near-zone multipoles.}
\begin{equation}\label{eq:Utail}
  \begin{aligned}
    \mathrm{U}^{\text{tail}}_L(U) & = 2 G E \int_0^{+\infty}\!\! d \tau\, I_L^{(\ell+2)}(U-\tau)\left[\log \left(\frac{\tau}{2 b_0}\right)+\kappa_{\ell}\right], \\
    \mathrm{V}^{\text{tail}}_L(U) & = 2 G E \int_0^{+\infty}\!\! d \tau\, J_L^{(\ell+2)}(U-\tau)\left[\log \left(\frac{\tau}{2 b_0}\right)+\pi_{\ell}\right],    
    \end{aligned}
\end{equation}
with
\begin{equation}
\begin{aligned}\label{eq:kappa}
  \kappa_{\ell} & =\frac{2 \ell^2+5 \ell+4}{\ell(\ell+1)(\ell+2)}+\sum_{k=1}^{\ell-2} \frac{1}{k} \\
  \kappa_2 & = \frac{11}{12}\;,\quad \kappa_3 = \frac{97}{60} \,,\quad \kappa_4 = \frac{59}{30}\,, \quad \ldots \quad .
\end{aligned}
\end{equation}
See~\cite{Blanchet:2013haa} for an explicit formula for the $\pi_\ell$ appearing in the $V_L$'s: we will not need it in our current analysis as the contributions depending on $\pi_\ell$ are negligible in the soft-limit including at the order $\omega\log\omega$. Of course we could change the values of $\kappa_\ell$ and $\pi_\ell$ by an overall shift if we redefine the cutoff $b_0$ appearing in~\eqref{eq:Utail}. Taking the Fourier transform of~\eqref{eq:Utail} to frequency space, we get 
\begin{equation}\label{eq:Utailom}
\mathrm{U}^{\text{tail}}_L(U)= 2 G E i \omega I_L^{(\ell)}(\omega)\Big[\log(2 b_0 \omega) - \gamma - \kappa_{\ell} - 
i \, \frac{\pi}{2}\Big]\,.
\end{equation}

Let us stress that the result \eqref{eq:Utail} is quite surprising from the amplitudes point of view: when evaluated at the first non-trivial order, ${\mathcal O}(G^2)$, it implies that the half-integer PN corrections to each multipole are proportional to the tree-level result (since at order $G$ we have $\mathrm{U}_L = I_L^{(\ell)}$). As discussed above, these contributions come from the $c$-channel cuts on the amplitudes side and there is no obvious reason why they should be related in a simple way to the tree-level result. Another important point is that, by taking the soft limit~\eqref{eq:Utailom}, there are two sources of $\omega\log\omega$ terms: the first one is the $(\omega\log\omega)^0$ contribution from the ${\mathcal O}(G)$ term of $I_L^{(\ell)}$  ({\rm i.e.} the tree-level multipole moment) and the factor of $\log\omega$ in the square parenthesis, while the second contribution comes from the constant terms in the same square parenthesis multiplied by the $\omega^0 \log\omega$ terms from $I_L^{(\ell)}$ (note the overall $\omega$ in \eqref{eq:Utailom}). Thus, at this order, the soft result is sensitive to values of $\kappa_\ell$ reported in~\eqref{eq:kappa} and, as already mentioned in the previous section, to non-universal parts of the soft waveform. We can derive this second, non-universal, contribution by using the $\omega^0$ term in \eqref{omega0} and the first one, which depends on the $\kappa_\ell$'s, by using the $\log\omega$ term in \eqref{logomega}. The terms with an extra factor of $\pi$ simply reproduce the PN-odd terms of the first contribution in~\eqref{eq:W1pinopi}, as is clear by comparing \eqref{eq:B1pi} with \eqref{eq:Utailom}. Therefore, adjusting the arbitrary scales by
\begin{equation}\label{choiceb0}
		\log (2 b_0) = \log\tilde{b}_0+ \gamma\,,
\end{equation}
we arrive at the following result for the counterpart of $\tilde W_{C}^{[\omega\log\omega]}$ in \eqref{eq:W1pinopi} as obtained from the MPM approach~\cite{Blanchet:2013haa},
\begin{align}
	\label{MPMomegalogomegafull}
		&\frac{\tilde W_{C}^{[\omega\log\omega],\text{MPM}}}{\kappa \omega\log\omega}
		\\
		\nonumber
		&=
		-\frac{1}{24\,p_\infty} i G^2 m^3 \nu  \left(35 (\cos (2 \theta )+3) \cos (2 \phi )+140 i \cos (\theta ) \sin (2 \phi )+22 \sin ^2(\theta )\right)
		\\
		\nonumber
		&+
		\frac{1}{60} G^2 m^3 \Delta\, \nu  \sin (\theta ) (\cos (\phi )+i \cos (\theta ) \sin (\phi )) (\cos (\theta ) (307 \cos (2 \phi
		)-67)+614 i \sin (\phi ) \cos (\phi ))
		\\
		\nonumber
		&+\frac{1}{120}\, p_\infty i G^2 m^3 \nu  \Big[
		\Big(79 \cos (2 \theta )+57 \cos (4 \theta )+420 \Big) \cos (2 \phi )
		\\
		\nonumber
		&+2 \sin ^2(\theta ) \Big(51 (\cos (2
		\theta )+3) \cos (4 \phi )+204 i \cos (\theta ) \sin (4 \phi )+63 \cos (2 \theta )+400\Big)
		\\
		\nonumber
		&+4 i \cos (\theta ) (42 \cos (2 \theta )+97) \sin (2
		\phi )
		\Big]
		+\mathcal O(p_\infty^2)
		+
		(\text{quadratic in }\nu)\,.
\end{align}
For $\theta= \frac{\pi}{2}$, we have checked that \eqref{MPMomegalogomegafull} agrees with the corresponding terms of the MPM result in Eq.~(9.11) of \cite{Bini:2023fiz}.
By comparing~(\ref{MPMomegalogomegafull}) and~(\ref{PNomegalogomegafull}) we note that, at first sight, they differ by an $\mathcal O(p_\infty)$ term.
However, we can find agreement also at that order 
by performing the following shift in the retarded time $U$ appearing in \eqref{WaveformPMfreqintegral0},
	\begin{equation}\label{Ufixconst}
		U \mapsto U + 2G m p_\infty^2
	\end{equation}
and expanding for small $G$. This is equivalent to considering the infinitesimal (retarded) time translation of the metric fluctuation
\begin{equation}\label{}
	h_{\mu\nu} = g_{\mu\nu}-\eta_{\mu\nu}\,,\qquad
	\delta h_{\mu\nu} 
	= 
	2G m p_\infty^2\,\partial_U h_{\mu\nu}\,,
\end{equation}
or to deforming the cutoff redefinition \eqref{choicemuIR} in a manner similar to \eqref{Ufixconst}. The order $\nu$ part of the waveform can be obtained also by studying the perturbation around the solution describing a single black hole, see for instance~\cite{Mino:1997bx}. This approach has been reconsidered in~\cite{Fucito:2023afe} by using recent progress on the connection problem for the Heun equation and it would be interesting to compare also with the results obtained in this way.

Let us now move on to the $\mathcal O(G^2\nu^2)$ terms, for which, using the soft-expanded amplitude results of Section~\ref{sec:Soft}, we find 
\begin{equation}\label{Wordinenuquadro}
	\begin{split}
		&\frac{\tilde W_{C}^{[\omega\log\omega]}}{\kappa \omega\log\omega}
		=
		(\text{linear in }\nu)
		\\
		&
		+
		\frac{G^2 m^3 \nu^2}{160} \,p_\infty\, \Big[822 \sin (2 \theta ) \sin (\theta ) \sin (4 \phi )-i (142 \cos (2 \theta )+231 (\cos (4 \theta )+5)) \cos (2 \phi )\\
		&-i \sin ^2(\theta ) (\cos
		(2 \theta ) (411 \cos (4 \phi )+513)+1233 \cos (4 \phi )+775)\\
		&+4 \cos (\theta ) (171 \cos (2 \theta )+211) \sin (2 \phi )\Big]
		+\mathcal O(p_\infty^2)\,.
	\end{split}
\end{equation}
On the MPM side, in addition to contributions arising from the tails, $\nu^2$ terms also receive nonlinear contributions arising from relations among near- and intermediate-zone multipoles that need to be combined in order to produce the far-zone ones and start appearing at relative $1/c^5$ PN order. For example, from \cite{Bini:2023fiz},
\begin{equation}\label{Uijklnonlinear}
	U_{ijkl}^{QQ}
	=
	G
	\left[
	-\frac{21}{5}\, I_{\langle ij}^{\phantom{(5)}} I_{kl\rangle}^{(5)}
	-\frac{63}{5}\, I_{\langle ij}^{(1)} I_{kl\rangle}^{(4)}
	- \frac{102}{5}\,I_{\langle ij}^{(2)} I_{kl\rangle}^{(3)}
	\right].
\end{equation}
Note that contributions of this kind start mattering in $\mathcal O(G^2)$ multipoles, and hence in the $\mathcal O(G^3)$ metric fluctuation, because the quadrupole moment employed in the PN literature has a $G$-independent piece that is determined by the free trajectories and is quadratic in the time $t$. As a result, $I_{ij}$, $I_{ij}^{(1)}$ and $I_{ij}^{(2)}$ start at $\mathcal O(G^0)$, while $I_{ij}^{(\ge 3)}$ start at order $G$. Moreover, contributions like \eqref{Uijklnonlinear} are clearly quadratic in $\nu$, since each multipole brings about a factor of the mass-ratio.
Combining the $\nu^2$ arising from the tail formula \eqref{eq:Utail} with the quadratic contributions akin to \eqref{Uijklnonlinear}, we obtain
\begin{align}\label{Unu2MPM}
	&
	\frac{\tilde W_{C}^{[\omega\log\omega],\text{MPM}}}{\kappa \omega\log\omega}
	=
	(\text{linear in }\nu)\\
	\nonumber
	&
	-\frac{G^2 m^3 \nu^2}{160}\,i\,p_\infty  \Big[
	842 i \sin (2 \theta ) \sin (\theta ) \sin (4 \phi )+(102 \cos (2 \theta )+241 \cos (4 \theta )+1025) \cos (2
	\phi )\\
	\nonumber
	&+\sin ^2(\theta ) (421 (\cos (2 \theta )+3) \cos (4 \phi )+543 \cos (2 \theta )+705)
	\\
	\nonumber
	&+4 i \cos (\theta ) (181 \cos (2 \theta )+161)
	\sin (2 \phi )
	\Big]+\mathcal O(p_\infty^2)\,.
\end{align}
We have cross-checked  \eqref{MPMomegalogomegafull} against the corresponding terms in Eq.~(9.11) of \cite{Bini:2023fiz} for $\theta= \frac{\pi}{2}$.
Once again, the two results \eqref{Unu2MPM} and \eqref{Wordinenuquadro} do not match at face value, but we can find agreement upon performing the following transformation in \eqref{WaveformPMfreqintegral0},
\begin{equation}\label{UfixST}
	U \mapsto U - T(n)\,,\qquad T(n) = 2 G (m_1\alpha_1\log\alpha_1+m_2\alpha_2\log\alpha_2)
\end{equation}
and expand for small $G$. This is equivalent to an infinitesimal transformation,
\begin{equation}\label{}
	\delta_T h_{\mu\nu} = - T(n)\,\partial_U h_{\mu\nu}\,.
\end{equation}
Up to static contributions, which we systematically disregard by focusing on $\omega>0$ in frequency domain (and which would only affect the $\mathcal O(G)$ metric fluctuation), we recognize this as the action of the supertranslation 
(see e.g.~\cite{Sachs:1962zza,Barnich:2010eb,He:2014laa,Strominger:2014pwa})
\begin{equation}\label{supertranslation}
	\delta_T h_{AB} = - T(n) \,\partial_u h_{AB} + r \left[
	2 D_A D_B - \gamma_{AB} \Delta
	\right]T(n)
\end{equation}
on the angular part of the metric, $h_{AB} = r^2\,(\partial_A \hat n)^i\,(\partial_B \hat n)^j h_{ij}$ where $A,B=\theta,\phi$, while $\gamma_{AB}=(\partial_A \hat n)\cdot(\partial_B \hat n)$ is the metric on the $2$-sphere, $D_A$ is the associated covariant derivative and $\Delta = \gamma^{AB}D_A D_B$.
Following the terminology of Ref.~\cite{Veneziano:2022zwh}, the supertranslation appearing in \eqref{supertranslation} is precisely the one needed to move from the ``canonical'' BMS frame, where the asymptotic shear vanishes in the far past, to the ``intrinsic'' one, where the asymptotic shear is nonzero even in the far past and in the PM expansion starts with an $\mathcal O(G)$ constant term dictated by the velocities of the incoming objects.
The intrinsic frame is the standard supertranslation frame employed in the PN literature, where the static terms in the leading-order multipoles play a crucial role, for instance via non-linearities such as Eq.~\eqref{Uijklnonlinear}.

The combined effect of \eqref{Ufixconst}, \eqref{UfixST} is to fully resolve the following ``mismatch''
between the $\omega\log\omega$ waveform the amplitude-based approach, given by \eqref{PNomegalogomegafull}, \eqref{Wordinenuquadro}, 
and the analogous MPM result, given by \eqref{MPMomegalogomegafull}, \eqref{Unu2MPM}, 
\begin{equation}\label{mismatchgenericth}
	\begin{split}
		&\frac{\tilde W_{C}^{[\omega\log\omega]}-\tilde W_{C}^{[\omega\log\omega],\text{MPM}}}{\kappa\omega\log\omega}
		\\
		&=
		-i G^2 m^3 \nu  p_\infty \left(\nu  \sin ^2(\theta ) \sin ^2(\phi )+\nu -2\right) (\cos (\phi )+i \cos (\theta ) \sin (\phi ))^2 + \mathcal O(p_\infty^2)\,.
	\end{split}
\end{equation}
Restricting to the equatorial plane $\theta=\frac{\pi}{2}$,  and focusing on terms that are even under $\phi\to\phi+\pi$, Eq.~\eqref{mismatchgenericth} reduces to
\begin{equation}\label{eq:mismst}
	\begin{split}
&\frac{\tilde W_{C}^{[\omega\log\omega]}-\tilde W_{C}^{[\omega\log\omega],\text{MPM}}}{\kappa\omega\log\omega} \Big|_{\theta=\frac{\pi}{2}} \\
&= \frac{1}{2}\,p_\infty\, i G^2 m^3 \nu  \cos ^2(\phi ) (4+\nu(\cos(2\phi)-3))+\mathcal O(p_\infty^2) + (\text{odd})\,,
	\end{split}
\end{equation}
in agreement with the difference between
\eqref{PNomegalogomegafull}, \eqref{Wordinenuquadro} and Eq.~(9.11) of Ref.~\cite{Bini:2023fiz}.
The result in \eqref{eq:mismst} has also been derived by~\cite{toapFei}. 
To reiterate, with the supertranslation function $T(n)$  given in \eqref{UfixST} \cite{Veneziano:2022zwh}, one can obtain agreement, between the MPM expression of Ref.~\cite{Bini:2023fiz} and the waveform obtained from the amplitude computed in \cite{Brandhuber:2023hhy,Herderschee:2023fxh,Georgoudis:2023lgf} in the PN limit,
\begin{equation}\label{}
	\frac{\tilde W_{C}^{[\omega\log\omega]}-\tilde W_{C}^{[\omega\log\omega],\text{MPM}}-i \omega\,(2 G m p_\infty^2-T(n))\tilde W_0^{[\log\omega]}}{\kappa\omega\log\omega}=\mathcal O(p_\infty^2)\,.
\end{equation}

\section{Conclusions and Outlook}
\label{sec:outlook}

In this paper, we investigated the gravitational scattering waveform at subleading order, shedding light on its structure and properties. 
We first clarified the role played by the KMOC cut contribution involving intermediate states with two massive particles pointed out in Ref.~\cite{Caron-Huot:2023vxl}, and which had been neglected in Refs.~\cite{Brandhuber:2023hhy,Herderschee:2023fxh,Elkhidir:2023dco,Georgoudis:2023lgf}. We have shown that it can be neatly rewritten as a simple differential operator acting on the tree-level waveform, as in Eq.~\eqref{CHsolved}.  Its inclusion thus simply amounts to rotating the basis vectors $b_J\mapsto b_e$, $-p_{1,2}\mapsto \tilde p_{1,2}$ as in \eqref{barp1barp2}, \eqref{betransf}, much like it happens for the subleading PM impulse, and shifting the origin of (retarded) time. This  translation in time is in fact divergent and is related to the Shapiro time delay, as already recognized in \cite{Caron-Huot:2023vxl,Bini:2023fiz}.
This result clarifies how amplitude-based calculations of observables, in particular both the impulse and the waveform, calculated in the KMOC approach consistently hold in the $b_J$, $-p_{1,2}$ frame. Instead the eikonal approach works in the $b_e$, $\tilde p_{1,2}$ frame, so that the cut pointed out in \cite{Caron-Huot:2023vxl} does not appear, and the naive exponentiation of~\cite{Cristofoli:2021jas} seems to work also at this order. The eikonal approach yields directly the frame used in MPM approach.

Our next target has been the expansion of the waveform in the soft limit $\omega\to0$, for generic velocities. Besides checking that the $\frac{1}{\omega}$, $\log\omega$, $\omega(\log\omega)^2$ contributions in this limit perfectly match the universal predictions dictated by soft theorems \cite{Saha:2019tub,Sahoo:2021ctw}, we have also been able to obtain a compact explicit expression, detailed in Eq.~\eqref{omegalogomega}, \eqref{omega0}, \eqref{calMomegalogomega} for the first non-universal contribution, $\omega\log\omega$. This is not fixed by soft theorems at one-loop, \emph{i.e.}~subleading PM, level. For this calculation, we developed a method of expansion by regions adapted to the soft limit of the waveform.

Using such explicit and handy results, we have been able to revisit the mismatch discussed in Ref.~\cite{Bini:2023fiz} between the waveform obtained from the amplitude EFT approach and the one derived from the MPM formalism in the PN limit, \emph{i.e.}~for small velocities. While all universal soft terms match straightforwardly between the two approaches, the comparison for $\omega\log\omega$ requires more care. Still, we have shown that the agreement between the amplitude and MPM results can be recovered by suitably shifting the origin of retarded time and by performing the BMS supertranslation \eqref{supertranslation}. The latter connects the canonical ``asymptotic'' frame~\cite{Veneziano:2022zwh}, where the early-time shear is zero and where the amplitude result holds, to the ``intrinsic'' one, where the shear has a constant $\mathcal O(G)$ piece as systematically assumed in the PN literature. This solves the mismatch pointed out in \cite{Bini:2023fiz} up to and including relative 2.5PN \emph{i.e.}~$1/c^5$ order.

The presence of such supertranslation-dependent static terms is indeed crucial in the MPM-PN formalism, where several $\mathcal O(G^2)$ multipoles take the form of non-linear contributions given by $G$ times an $\mathcal O(G^0)$ static piece times an $\mathcal O(G)$ dynamic piece.
In other words, such terms arise due to the non-linearities of gravity. On the contrary, the amplitudes approach discussed here should be interpreted ``without the static modes'' to use a nomenclature introduced in~\cite{DiVecchia:2023frv}. One should then perform the supertranslation derived in~\cite{Veneziano:2022zwh} to compare with the PN results. However, when working at $\mathcal O(G^3)$ with the full-time dependent waveform, the action of the supertranslation involving a time-derivative in Eqs.~\eqref{UfixST}, \eqref{supertranslation}, and not just on the static $\mathcal O(G)$ part, must be taken into account. It would be interesting to explore if the amplitude approach can be adapted to produce the MPM result directly in the supertranslation frame ``including the static modes'', as well as to study the dependence on the asymptotic frame in the presence of additional vector or scalar fields.

Let us now move to the open issues.
It would be of course interesting to go beyond the soft limit and check the mechanisms proposed here on the full PN waveform up to $1/c^5$ so as to complete the comparison with~\cite{Bini:2023fiz}. 
In this respect, it will be important to better understand why we obtain a different result for the difference between amplitude and MPM waveform compared to Eq.~(9.13) of~\cite{Bini:2023fiz}. In that reference, the amplitude-based result was obtained by doing the expansions in the opposite order compared to the approach we took here, {\rm i.e.}~starting with low velocity limit and then performing the soft limit.
We expect that the comparison between amplitude and MPM results for the subleading PM waveform will work to any PN order and any frequency up to properly adjusting the (super)translation frame as discussed here. Of course, it would be very interesting to perform such comparison explicitly beyond what has it has been considered in this paper: this would provide highly non-trivial checks both on the expressions entering the quasi-Keplerian orbits and on the various relations that link, within the MPM approach, the near-zone field with the asymptotic values  at null infinity (see for instance~\cite{Mishra:2015bqa,Cho:2018upo,Bini:2022enm} and references therein).

A systematic expansion of the subleading PM waveform in the PN limit calls for a rewriting of the 5-point 1-loop amplitude in order to tame the technical complication arising from the presence of spurious poles in its expression. The issue here is that, while the full result for the waveform is well-behaved in the PN limit, separately its different contributions involving rational as well as logarithmic terms are very singular. The smooth PN limit is recovered only thanks to highly nontrivial cancellations among all various pieces. This mechanism is already present, in a harmless form, e.g.~in the results for the radiated energy-momentum \cite{Herrmann:2021lqe}, but given the much larger size of the waveform expression, the presence of such singularities in intermediate results represents a nontrivial technical obstacle for performing the PN (and, partly, also soft) expansion of this observable. It would thus be highly desirable to reorganize the result in a way, if any, that makes such spurious poles harmless (and steps in this directions have already been taken~\cite{Bohnenblust:2023qmy}).

Another potential development concerns the extension to NLO of waveform calculations to take into account physical effects associated to more complicated binaries, for instance involving spinning black holes or different astrophysical objects, like neutron stars, that can be subject to tidal deformations. It will be interesting to test the role of the supertranslation frame also in those contexts, to see whether the pattern found here is, as we suspect, more general.

Moving forward, it will be important to put to the test the straightforward exponentiation pattern dictated by the eikonal operator proposed in \cite{Cristofoli:2021jas,DiVecchia:2022piu}, for which the next nontrivial check is represented by the classical 2-loop 5-point amplitude.
The present work highlights an additional piece of appeal of that calculation, which would make it possible to check non-linearities involving all non zero-frequency gravitons.  

\acknowledgments
We would like thank Francesco Alessio, Donato Bini, Lara Bohnenblust, Thibault Damour, Paolo Di Vecchia, Andrea Geralico, Radu Roiban, Fei Teng, Gabriele Veneziano for several interesting conversations and suggestions, and Stefano De Angelis for collaboration on a related project. 
A.~G.~is supported by a Royal Society funding, URF\textbackslash{R}\textbackslash221015.
C.~H.~is supported by UK Research and Innovation (UKRI) under the UK government’s Horizon Europe funding guarantee [grant EP/X037312/1 “EikoGrav: Eikonal Exponentiation and Gravitational Waves”]. R.~R. is partially supported by the UK EPSRC grant ``CFT and Gravity: Heavy States and Black Holes'' EP/W019663/1 and the STFC grants "Amplitudes, Strings and Duality", grant numbers ST/T000686/1 and ST/X00063X/1. R.~R. would like to thank IHES for hospitality during the initial stages of this project. No new data were generated or analysed during this study.

\providecommand{\href}[2]{#2}\begingroup\raggedright\endgroup

\end{document}